\documentclass[11pt]{article}
\usepackage{amsfonts}
\usepackage{amssymb}
\usepackage{latexsym}
\usepackage{amsmath}
\usepackage{graphicx}
\usepackage{float}

\numberwithin{equation}{section}

\newcommand{\be}{\begin{equation}}
\newcommand{\ee}{\end{equation}}
\newcommand{\bes}{\begin{equation*}}
\newcommand{\ees}{\end{equation*}}
\newcommand{\eqn}{\begin{eqnarray}}
\newcommand{\feqn}{\end{eqnarray}}
\newcommand{\eqnn}{\begin{eqnarray*}}
\newcommand{\feqnn}{\end{eqnarray*}}

\begin{document}

\begin{titlepage}

\begin{flushright}
CERN-PH-TH/2012-305
\end{flushright}

\vskip 2.0 cm
\begin{center}  {\huge{\bf      Multi-Centered Invariants,\\\vskip 0.5 cm Plethysm and Grassmannians}}

\vskip 1.5 cm

{\Large{\bf Sergio L. Cacciatori$^{1}$}, {\bf Alessio Marrani$^2$}, {\bf Bert van Geemen$^3$}}

\vskip 1.5 cm

$^1${\sl Dipartimento di Scienze ed Alta Tecnologia,\\Universit\`a degli Studi dell'Insubria, Via Valleggio 11, I-22100 Como, Italy\\
and\\ INFN, Sezione di Milano, Via Celoria 16, 20133 Milano, Italy\\
\texttt{sergio.cacciatori@uninsubria.it}}\\

\vskip 0.5 cm

$^2${\sl Physics Department,Theory Unit, CERN,\\CH 1211, Geneva 23, Switzerland\\
\texttt{alessio.marrani@cern.ch}}\\

\vskip 0.5 cm

$^3${\sl Dipartimento di Matematica, Universit\`a di Milano,\\Via Saldini 50, I-20133 Milano, Italy\\
\texttt{lambertus.vangeemen@unimi.it}}\\

 \end{center}

 \vskip 2.0 cm

\begin{abstract}

Motivated by multi-centered black hole solutions of Maxwell-Einstein
theories of (super)gravity in $D=4$ space-time dimensions, we develop some
general methods, that can be used to determine all homogeneous invariant
polynomials on the irreducible $\left( SL_{h}(p,\mathbb{R}%
)\otimes G_{4}\right) $-representation $\left( \mathbf{p},R\right) $,
where $p$ denotes the number of centers, and $SL_{h}(p,\mathbb{R})$ is the
\textit{\textquotedblleft horizontal"} symmetry of the system, acting upon
the indices labelling the centers. The black hole electric and magnetic
charges sit in the symplectic representation $R$ of the generalized
electric-magnetic ($U$-)duality group $G_{4}$.

We start with an algebraic approach based on \textit{classical invariant
theory}, using Schur polynomials and the Cauchy formula. Then, we perform a
geometric analysis, involving Grassmannians, Pl\"{u}cker coordinates, and
exploiting Bott's Theorem.

We focus on \textit{non-degenerate} groups $G_{4}$ \textit{\textquotedblleft
of type }$E_{7}$\textit{"} relevant for (super)gravities whose (vector
multiplets') scalar manifold is a \textit{symmetric} space. In the
 \textit{triality-symmetric} $stu$ model of $\mathcal{N}=2$ supergravity, we explicitly construct a basis for the $10$
linearly independent degree-$12$ invariant polynomials of $3$-centered
black holes.

 \end{abstract}
\vspace{24pt} \end{titlepage}


\newpage \tableofcontents \newpage



\section{\label{Intro}Introduction}

The \textit{Attractor Mechanism} \cite{AM-Refs,FGK}, originally discovered
in $\mathcal{N}=2$, $D=4$ Maxwell-Einstein supergravity and then
investigated in other extended supergravities as well as in
non-supersymmetric theories of gravity (see \textit{e.g.} \cite{AM-Revs} for
reviews and list of Refs.), plays a central role in the physics of extremal
black holes (BHs), as well as of (intersecting configurations of) extremal
black $p$-branes \cite{Inters-Attr}, also in $D>4$ space-time dimensions. In
its simplest framework, namely in presence of Abelian vectors and scalar
fields in the background of an extremal BH, the area of the event horizon
can be expressed purely in terms of the fluxes of the $2$-form Abelian field
strengths and of their duals, whose fluxes define the magnetic and electric
BH charges, fitting a symplectic vector $Q$. The dynamics of the scalar
fields exhibits an \textit{attractor} phenomenon, namely the value of the
field at the BH event horizon is completely determined in terms of the
magnetic and electric charges, \textit{regardless} of the initial (boundary)
conditions defined for the flow at spatial infinity\footnote{%
Some exception/violations of the \textit{Attractor Mechanism} include
\textit{e.g.} the existence of \textit{basins of attraction}/\textit{area
codes} \cite{Area-Codes-Refs} as well as of \textit{moduli spaces}/\textit{%
flat directions} of attractor flows \cite{FM-2,Inters-Attr}.}. In general,
the near-horizon attractor dynamics can be reformulated in terms of critical
points of a \textit{BH effective potential} \cite{FGK}, which in presence of
an underlying local supersymmetry also enjoys a geometric interpretation in
terms of \textit{central charge}(s) and \textit{matter charges} (if any).

The vector space of electric-magnetic BH charges generally defines an
irreducible\footnote{%
This strictly holds for \textit{unified} theories, in which \textit{all}
Abelian $2$-form field strengths (and their duals) transform in an irrep. of
$G_{4}$; the following reasoning can be easily generalized to\textit{\
non-unified} frameworks.} representation (irrep.) space $R$ for the
generalized electric-magnetic ($U$-) duality\footnote{%
Here $U$-duality is referred to as the \textquotedblleft
continuous\textquotedblright\ symmetries of \cite{CJ-1}. Their discrete
versions are the $U$-duality non-perturbative string theory symmetries
introduced by Hull and Townsend \cite{HT-1}.} group $G_{4}$. Under the
action of $G_{4}$, the irrep. space $R$ undergoes a \textit{stratification}
into orbits, which in turn are in correspondence with classes of BHs, with
both regular and vanishing near-horizon geometry (corresponding to \textit{%
\textquotedblleft large"} and \textit{\textquotedblleft small"} BHs,
respectively); thus, the classification of $G_{4}$-orbits in $R$ results in
a group-theoretical characterization of BH solutions themselves. In
Maxwell-Einstein supergravity theories whose scalar manifold is a \textit{%
symmetric} space $G_{4}/H_{4}$ (with $H_{4}$ being the maximal compact
subgroup of $G_{4}$), the classification of orbits can be algebraically
achieved in terms of constraints imposed on the \textit{unique} \cite{Kac}
algebraically independent $G_{4}$-invariant homogeneous polynomial $\mathcal{%
I}$ in the irrep. $R$ (see \textit{e.g.} \cite{D=4-Orbits}, as well as \cite%
{LA10} for a recent \textit{r\'{e}sum\'{e}} and a list of Refs.).

Within this rather broad class of $D=4$ theories, $\mathcal{I}$ is a \textit{%
quadratic} polynomial ($\mathcal{I}=\mathcal{I}_{2}$) for $\mathcal{N}=2$
\textit{minimally coupled} \cite{Luciani,Gnecchi-1} as well as for $\mathcal{%
N}=3$ \cite{N=3} supergravity. In the remaining $D=4$ theories with \textit{%
symmetric} scalar manifolds, $G_{4}$ can be characterized (in terms of $R$)
as a group \textit{\textquotedblleft of type }$E_{7}$\textit{"} \cite%
{brown,Refs-Garibaldi,Duff-E7,FK-E7,FKM-E7,Stora-Proc}. In particular, the
charge representation $R$ satisfies
\begin{equation}
\dim \wedge ^{2}R=\dim S^{4}R=1.
\end{equation}%
Namely, the flux irrep. $R$ is \textit{symplectic} (\textit{i.e.}, endowed
with a unique symplectic structure $\mathbb{C}_{\left[ MN\right] }:=\mathbf{1%
}\in \wedge ^{2}R=:R_{a}^{\otimes 2}$, as it generally holds in $D=4$), and
it exhibits a unique, algebraically independent, degree-$4$ homogeneous
invariant polynomial\footnote{%
Actually, this characterizes $G_{4}$ (which can be simple or semi-simple) as
a \textit{non-degenerate} group \textquotedblleft of type $E_{7}$". The
\textit{\textquotedblleft degeneration"} of the $U$-duality symmetry in some
$\mathcal{N}=2$ theories \cite{Luciani,Gnecchi-1} and in $\mathcal{N}=3$
\cite{N=3} supergravity, and its relation to the \textit{minimal coupling}
of vector and scalar fields in Maxwell-Einstein (super)gravity theories in $%
D=4$ has recently been investigated in \cite{FKM-E7}.} $\mathcal{I}=\mathcal{%
I}_{4}$, related to a rank-$4$ completely symmetric $G_{4}$-invariant tensor
(the so-called $K$-tensor \cite{Exc-Reds,ADFMT-1,CFMY-Small-1,FMY-CV-1}) $%
K_{\left( MNPQ\right) }:=\mathbf{1}\in S^{4}R=:R_{s}^{\otimes 4}$. Simple
and semi-simple \textit{non-degenerate} $U$-duality groups $G_{4}$
\textquotedblleft of type $E_{7}$" relevant to the class of $D=4$
Maxwell-Einstein (super)gravity theories under consideration are listed in
Table 1 at the start of Sec.\ \ref{Examples}.

The properties of the quartic polynomial $\mathcal{I}_{4}$ constructed from
the $K$-tensor have been exploited in order to characterize in an algebraic
way the various scalar flows in the background of extremal \textit{%
single-centered} BHs \cite{D=4-Orbits}. The classification can be extended
to \textit{multi-centered} BHs \cite%
{Denef,Multi-ctr-Refs,almost-BPS,Bena-1,Bossard-Ruef,Bossard-Octonionic,Yeranyan-T^3}%
. In the case of $2$-centered solutions, a group theoretical study of the
invariant structures which can be defined in the vector space of
electric-magnetic fluxes has been started in \cite{FMOSY-1}, and then
developed in \cite{ADFMT-1,CFMY-Small-1,FMY-CV-1,Stora-Proc}; the connection
between $2$-centered invariant structures for the so-called $stu$ model \cite%
{STU} of $\mathcal{N}=2$, $D=4$ supergravity and \textit{Quantum Information
Theory} has then been investigated in \cite{Levay-2-ctr}. Furthermore,
relations between the $K$-tensor of the $stu$ model (giving rise to the
so-called \textit{Cayley's hyperdeterminant} \cite{Cayley,Duff-Cayley}) and
elliptic curves has been recently studied in \cite{Gibbs}, and extended to
the $2$-centered case in \cite{Levay-2-ctr}.

Besides the importance of the \textit{symplectic product} $\mathcal{W}$ (see
Eq.\ (\ref{W}) below) in order to define \textit{mutually non-local} charge
vectors pertaining to different centers \cite{Denef}, the physical relevance
of some higher-order $U$-invariant polynomials has been suggested in recent
investigations \cite{Bossard-Octonionic}, and further study in such a
direction is surely deserved in order to unravel their role \textit{e.g.} in
the spatial structure of general stationary \textit{almost-BPS} \cite%
{almost-BPS,Bossard-Ruef,Yeranyan-T^3} and \textit{composite non-BPS} \cite%
{Bossard-Ruef,Bossard-Octonionic,Yeranyan-T^3} multi-centered BH flows, with
\textit{flat} $D=3$ spatial slices as well as \textit{non-flat} ones \cite%
{Bena-1,Bena-2,Dall'Agata-U}.

In the case of BH solutions with $p$ centers, the $U$-duality group $G_{4}$
acts on $p$ copies of $R$; correspondingly, the charge vectors $Q^{a}$ carry
an index referring to the relevant center ($a=1,...,p$), and one has to
consider polynomial invariants in the $p\dim R$ coordinates on $R^{p}$.
Thus, a \textit{\textquotedblleft horizontal"} symmetry\footnote{%
The subscript \textquotedblleft $h$" stands for \textit{\textquotedblleft
horizontal"} throughout.} $SL_{h}(p,\mathbb{R})$, commuting with $G_{4}$,
naturally occurs. This was firstly introduced in \cite{FMOSY-1}, and it acts
on the index labelling the various centers, in such a way that $G_{4}$%
-invariant polynomials generally decompose into $SL_{h}(p,\mathbb{R})$%
-irreps. In the $2$-centered case ($p=2$), as mentioned, the problem of
determining a complete basis for the ring of $\left( SL_{h}(p,\mathbb{R}%
)\times G_{4}\right) $-invariant homogeneous polynomials has been solved in
\cite{FMOSY-1} and \cite{ADFMT-1}, respectively for semi-simple and simple
non-degenerate groups \textquotedblleft of type $E_{7}$" occurring as $U$%
-duality groups in $D=4$ supergravities with symmetric scalar manifolds%
\footnote{%
The same problem was solved, for a generic number $p$ of centers, in \cite%
{FMO-MS-1,FMY-CV-1} for simple, \textit{degenerate} groups \textquotedblleft
of type $E_{7}$" occurring in $\mathcal{N}=2$ \textit{minimally coupled} as
well as in $\mathcal{N}=3$ supergravity in $D=4$.}. Actually, the same
results had been obtained, within a completely different approach based on
nilpotent orbits, by Kac many years ago in \cite{Kac}; therein, it was also
shown that the complete basis composed by polynomials whose homogeneity
degree is the lowest possible is also \textit{finitely generating}, namely
all other higher-order invariant polynomials are simply polynomials in the
elements of the basis.

For example, in the $2$-centered \textit{simple} case \cite{Kac,ADFMT-1}
there are $7$ algebraically independent $U$-invariant polynomials, which
form a minimal degree complete basis for the corresponding ring; out of
them, $5$ are homogeneous of degree $4$ and they are arranged into a $%
\mathbf{5}$ (spin $s=2$) irrep. of the $2$-centered \textquotedblleft
horizontal" symmetry $SL_{h}(2,\mathbb{R})$, while the remaining ones are
polynomials homogeneous of degree $2$ and $6$ that are $SL_{h}(2,\mathbb{R})$%
-invariant (the one of degree $2$ is nothing but the \textit{symplectic
product} $\mathcal{W}$ defined in (\ref{W}) below). Out of these $7$ $G_{4}$%
-invariants, one can construct $4$ algebraically independent $\left(
SL_{h}(2,\mathbb{R})\times G_{4}\right) $-invariant polynomials, homogeneous
of degree $2$, $6$, $8$ and $12$ \cite{Kac,ADFMT-1}. With some abuse of
language, $\left( SL_{h}(p,\mathbb{R})\times G_{4}\right) $-invariants have
been usually named \textquotedblleft horizontal" invariants.

In the ($2$-centered) \textit{semi-simple} case \cite{Kac,FMOSY-1,FMY-CV-1},
further lower-order horizontal invariant structures arise as a consequence
of the factorization of the $U$-duality symmetry $G_{4}$; a particular,
noteworthy example is provided by the aforementioned $stu$ model, exhibiting
a \textit{triality symmetry }\cite{STU}, which should be modded out in order
to obtain invariant structures relevant for BHs (\textit{cfr.} the treatment
of \cite{FMOSY-1} \textit{vs.} \cite{Levay-2-ctr}, as well as the treatment
in\ Secs.\ \ref{STU-gen} and \ref{co10in}).\medskip

Although some general properties can be inferred from elementary group
theoretical considerations, a systematic study and classification of $\left(
p>2\right) $-centered solutions in terms of $\left( SL_{h}(p,\mathbb{R}%
)\times G_{4}\right) $-orbits is still lacking.

The aim of the present paper is to start developing some general methods
that can be used to determine all invariants associated to $p$-centered BH
solutions, for a generic $p$. In particular, we will be interested in $p$%
-centered horizontal invariants, namely homogeneous $\left( SL_{h}(p,\mathbb{%
R})\times G_{4}\right) $-invariant polynomials on the irrep. $\mathbb{R}%
^{p}\otimes R=:\left( \mathbf{p},R\right) $ of the overall symmetry $%
SL_{h}(p,\mathbb{R})\times G_{4}$ itself. The invariant polynomials
homogeneous of degree $k$ are clearly related to the $\left( SL_{h}(p,%
\mathbb{R})\times G_{4}\right) $-invariant tensors in the $k$-th completely
symmetric\footnote{%
The subscript \textquotedblleft $s$" ("$a$") stands for \textit{symmetric} (%
\textit{antisymmetric}) throughout.} power $S^{k}(\mathbb{R}^{p}\otimes
R)=:\left( \mathbf{p},R\right) _{s}^{\otimes k}$. This allows for the
exploitation of the \textit{classical invariant theory} (for which we will
mainly refer to the book \cite{Procesi}).\medskip

The plan of the paper is as follows.

In Sec.\ \ref{algebraic} we use the representation theory of a product group
$\mathcal{G}\times G$ in order to determine the corresponding invariant
structures. We first recall some general facts about invariant theory and,
in particular, the characterization of the $\left( \mathcal{G}\times
G\right) $-invariants in the symmetric products $S^{k}(U\otimes V)$ of the
irreps. $U$ and $V$ of $\mathcal{G}$ and $G$, respectively. By applying
these methods to the case $\mathcal{G}=SL_{h}(p,\mathbb{R})$ and $G=G_{4}$
relevant to $p$-centered (BH) solutions in $D=4$ supergravity, we can then
count $\left( SL_{h}(p,\mathbb{R})\times G_{4}\right) $-invariants\footnote{%
Up to a certain order, fixed by the available computing power (see analysis
in\ Sec.\ \ref{Examples}).} for all relevant \textit{generic}, \textit{simple%
} cases.

Next, in Sec.\ \ref{geometric}, we present a geometric analysis of the
invariants. We show that in the $p$-centered case the invariants can be
determined by using the \textit{Grassmannian} $Gr(p,R)$ of $p$-planes in $R$%
. This Grassmannian is embedded in a projective space by its \textit{Pl\"{u}%
cker coordinates}, which are global sections of a line bundle $L$ on $%
Gr(p,R) $. For any positive integer $a$, the group $GL(R)$, and thus%
\footnote{%
As also recently discussed in \cite{FGT}, the \textit{maximal} (but
generally \textit{non-symmetric}) embedding $G_{4}\subset Sp(R)$ (which in
supergravity is named \textit{Gaillard-Zumino} \cite{GZ} embedding) can be
regarded as a consequence of the following \textit{Theorem} by Dynkin (Th.
1.5 of \cite{dynkin-2}, more recently discussed \textit{e.g.} in \cite%
{Lorente}) : every \textit{irreducible} group of unimodular linear
transformations of the $N$-dimensional complex space (namely, a group of
transformations which does not leave invariant a proper subspace of such a
space) is maximal either in $SL(N)$ (if the group does not have a bilinear
invariant), or in $Sp(N)$ (if it has a skew-symmetric bilinear invariant),
or in $O(N)$ (if it has a symmetric bilinear invariant). Exceptions to this
rule are listed in Table VII of \cite{Lorente}.}%
\begin{equation}
G_{4}\subset Sp\left( R\right) \subset SL\left( R\right) \subset GL\left(
R\right) ,
\end{equation}%
acts on the sections $\Gamma (Gr(p,R),L^{\otimes a})$. These sections are
homogeneous polynomials of degree $a$ in the Pl\"{u}cker coordinates. Our
geometric characterization of the $\left( SL_{h}(p,\mathbb{R})\times
G_{4}\right) $-invariant polynomials, in combination with Bott's theorem
\cite{Bott-Th}, shows that all these invariants are given by $\left(
SL_{h}(p,\mathbb{R})\times G_{4}\right) $-invariant sections. In particular,
the $\left( SL_{h}(p,\mathbb{R})\times G_{4}\right) $-invariant polynomials
are generated by homogeneous polynomials in the Pl\"{u}cker coordinates.

Finally, in Sec.\ \ref{stu}, we present an application of the methods
developed in\ Secs. \ref{algebraic} and \ref{geometric} : in the \textit{%
semi-simple}, \textit{triality-symmetric} $\mathcal{N}=2$, $D=4$ $stu$
model, we compute a basis for the $10$-dimensional vector space of $\left(
SL_{h}(3,\mathbb{R})\times SL(2,\mathbb{R})^{3}\right) $-invariant
polynomials homogeneous of degree $12$ for $3$-centered BHs; the physical
issue of invariance under the symmetric group $S_{3}$, implementing the
\textit{triality symmetry} acting on the three copies of $SL(2,\mathbb{R})$
in the $U$-duality group $G_{4}=SL(2,\mathbb{R})^{3}$, is considered in
Secs. \ref{STU-gen} and \ref{co10in}.


\section{\label{algebraic}Algebraic Approach}


\subsection{Invariant Theory}

In order to tackle the problem of determining the invariants associated to
multi-centered BH solutions, we will make use of the \textit{classical
invariant theory}. Let us first collect some basic facts on how to find
invariants in $U\otimes V$ for the action of the group $GL(U)\times GL(V)$;
as mentioned above, we will mainly refer to the book \cite{Procesi}, to
which we address the reader for further details and a list of Refs.

\subsubsection{The Schur Polynomials}

A partition $\lambda $ of an integer $m\in \mathbb{Z}_{>0}$, denoted as $%
\lambda \vdash m$, is a non-increasing sequence $\lambda =(p_{1},\ldots
,p_{N})$ of integers $p_{i}\in \mathbb{Z}_{\geq 0}$ such that $%
\sum_{i=}^{N}p_{i}=m$. The number of non-zero elements in $\lambda $ is
denoted by $ht(\lambda ):=n$, so $p_{i}=0$ for $i>n$.

The \textit{Schur polynomial} $S_{\lambda }$ in $N$ variables $x_{1},\ldots
,x_{N}$, where $N\geq n:=ht(\lambda )$, is the symmetric polynomial, with
integral coefficients, defined as the quotient (\cite{Procesi}, 2.3.2)
\begin{equation}
S_{\lambda }(x)\,:=\,\frac{A_{\lambda +\rho }\left( x\right) }{V(x)},
\end{equation}%
where the partition $\lambda +\rho $ is defined as $\lambda +\rho
:=(p_{1}+N-1,p_{2}+N-2,\ldots ,p_{N})$, and $A_{\lambda +\rho }\left(
x\right) $ and $V(x)$ (\textit{Vandermonde determinant}) are two
anti-symmetric polynomials in $x_{1},\ldots ,x_{N}$, respectively given by
\begin{equation}
A_{(m_{1},\ldots ,m_{N})}(x)\,:=\,\sum_{\sigma \in S_{N}}\epsilon _{\sigma
}x_{\sigma (1)}^{m_{1}}x_{\sigma (2)}^{m_{2}}\cdots x_{\sigma (N)}^{m_{N}},
\end{equation}%
where $S_{N}$ is the group of permutations of $N$ variables and $\epsilon
_{\sigma }$ is the permutation parity, and
\begin{equation}
V(x)\,:=\,\prod_{1\leq i<j\leq N}(x_{i}-x_{j}).
\end{equation}%
Note that $S_{\lambda }=0$ if $ht(\lambda )>N$.

As from Th. 1 in \cite{Procesi}, 2.3.2, the Schur polynomials $S_{\lambda }$
with $\lambda \vdash m$ and $ht(\lambda )\leq N$ are a basis of the
polynomials in $N$ variables which are homogeneous of degree $m$ and are
invariant under permutations of the variables $x_{1},...,x_{N}$. Examples
are provided by the elementary symmetric functions
\begin{eqnarray}
S_{1^{h}}\, &=&\,\sum_{1\leq i_{1}<\ldots <i_{h}\leq
N}x_{i_{1}}x_{i_{2}}\cdots x_{i_{h}},\qquad \lambda \,=\,1^{h}\,:=(%
\underbrace{1,\ldots ,1}_{h},0,\ldots ,0);  \label{S_1^h} \\
S_{k}\,\, &=&\,\sum_{1\leq i_{1}\leq \ldots \leq i_{k}\leq
N}x_{i_{1}}x_{i_{2}}\cdots x_{i_{k}},\qquad \lambda \,=\,k\,:=(k,\underbrace{%
0,\ldots ,0}_{N-1}),  \label{S_k}
\end{eqnarray}%
which differ only in the possibility to consider or not the same values for
at least a pair of indices in the string $i_{1},\ldots ,i_{h}$.


\subsubsection{Traces of $GL$-Representations}

Let $V$ be a vector space of dimension $N$ with basis $v_{1},\ldots ,v_{N}$,
and let $y:=(y_{1},\ldots ,y_{N})\in (\mathbb{C}^{\ast })^{N}$ act by $%
diag(y_{1},\ldots ,y_{N})$ on $V$.\newline
A partition $\lambda $ with $ht(\lambda )\leq N$ defines an irreducible
representation $S_{\lambda }(V)$ of $GL(V)$ which is a summand of $\otimes
^{m}V$ where $\lambda \vdash m$ (\cite{Procesi} 9.3.1, (3.1.3)). If $%
ht(\lambda )>N$, then $S_{\lambda }(V)=0$. Moreover, any irreducible
representation of $GL(V)$ is isomorphic to an $S_{\lambda }(V)$ for a unique
partition $\lambda $ with $ht(\lambda )\leq N$ (\cite{Procesi}, 9.8.1). The
trace of $diag(y_{1},\ldots ,y_{N})\in (\mathbb{C}^{\ast })^{N}$ (\textit{%
i.e.} the standard maximal torus of $GL(V)$) on the irreducible
representation $S_{\lambda }(V)$ is the Schur polynomial $S_{\lambda
}(y_{1},\ldots ,y_{N})$. The dimension of the representation associated to
the partition $(p_{1},\ldots ,p_{N})$ is (\cite{Procesi}, 9.6.2):
\begin{equation}
\dim \,S_{\lambda }(V)\,=\,\prod_{1\leq i<j\leq N}\frac{p_{i}-p_{j}+j-i}{j-i}%
,\qquad \lambda =(p_{1},\ldots ,p_{N}).  \label{dim-S-lambda}
\end{equation}

For instance, $\lambda :=1^{h}$ defines $S_{\lambda }(V):=\wedge ^{h}V$ (\ref%
{S_1^h}), the rank-$h$ completely antisymmetric tensor representation of $%
GL(V)$, which has dimension $\binom{N}{h}$; in particular, the partition $%
\lambda =1^{N}$ selects the one-dimensional determinant representation on $%
\wedge ^{N}V$ (realized by the Ricci-Levi-Civita symbol $\epsilon
_{i_{1}...i_{N}}$). \newline
Another example is provided by the partition $\lambda :=k:=(k,\underbrace{%
0,\ldots ,0}_{N-1})$, which defines $S_{\lambda }(V):=S^{k}V$ (\ref{S_k}),
the $k$-th symmetric product of $V$, namely the rank-$k$ completely
symmetric tensor representation of $GL(V)$. A basis of $S^{k}V$ is provided
by $v_{1}^{a_{1}}\cdots v_{N}^{a_{N}}$ with $a_{i}\geq 0$ and $%
\sum_{i=1}^{N}a_{i}=k$, and the action of $y$ on this basis elements is the
multiplication by $y_{1}^{a_{1}}\cdots y_{N}^{a_{N}}$. Hence, the trace of $%
y $ on $S^{k}V$ is the sum of all monomials in $y_{1},\ldots ,y_{N}$ which
are homogeneous of degree $k$. As mentioned, this is the Schur polynomial $%
S_{k}$ (\ref{S_k}), so $tr(y|S^{k}V)=S_{k}(y)$. A generating function for
these $S_{k}$ can be obtained by noting that%
\begin{eqnarray}
&&\left( 1+...+y^{a_{1}}t^{a_{1}}+...\right) \left(
1+...+y^{a_{2}}t^{a_{2}}+...\right) ...\left(
1+...+y^{a_{N}}t^{a_{N}}+...\right)  \notag \\
&=&1+S_{1}\left( y\right) t+S_{2}(y)t^{2}+...+S_{k}(y)t^{k}+...,
\end{eqnarray}%
and it is given by the \textit{Molien formula} ($S_{0}(y)=1$; \cite{Procesi}%
, 9.4.3, (4.4.3)):
\begin{equation}
\prod_{j=1}^{N}\frac{1}{1-y_{j}t}\,=\,\sum_{k=0}^{\infty }S_{k}(y)t^{k}.
\label{Molien}
\end{equation}


\subsubsection{\label{deccc}Decomposing $S^{k}(U\otimes V)$}

A generalization of the Molien formula (\ref{Molien}), which yields the
decomposition of $S^{k}(U\otimes V)$ under $GL(U)\times GL(V)$, is provided
by the following formula, due to Cauchy. Let\footnote{%
The formula (\ref{Cauchy}) is proven in \cite{Procesi}, 2.3.4 for $n=m$, but
setting $x_{i}=0$ for $m\leq i\leq n$, the proof holds for $m\leq n$.} $%
m\leq n$ be two positive integers, then:
\begin{equation}
\prod_{i=1}^{m}\prod_{j=1}^{n}\frac{1}{1-x_{i}y_{j}}\,=\,\sum_{\lambda
:ht(\lambda )\leq m\leq n}\,S_{\lambda }(x_{1},\ldots ,x_{m})S_{\lambda
}(y_{1},\ldots ,y_{n}).  \label{Cauchy}
\end{equation}%
The interpretation of the Cauchy formula (\ref{Cauchy}) in terms of
characters of representations is given \textit{e.g.} in \cite{Procesi},
9.6.3. Let $U$ and $V$ be vector spaces of dimension $m$ and $n$
respectively, and assume that $m\leq n$. Let $u_{1},\ldots ,u_{m}$, $%
v_{1},\ldots ,v_{n}$ be bases of $U$,$V$ respectively, and let $x\in (%
\mathbb{C}^{\ast })^{m}$, $y\in (\mathbb{C}^{\ast })^{n}$ act on these
spaces by $diag(x_{1},\ldots ,x_{m})$, $diag(y_{1},\ldots ,y_{n})$. The
eigenvalues of $(x,y)$ on $U\otimes V$ are then the $x_{i}y_{j}$ with $1\leq
i\leq m$ and $1\leq j\leq n$. Thus, Cauchy formula (\ref{Cauchy}) implies
that
\begin{equation}
\sum_{k=0}^{\infty }tr((x,y)|S^{k}(U\otimes V))t^{k}\,=\,\sum_{\lambda
:ht(\lambda )\leq m\leq n}S_{\lambda }(x)S_{\lambda }(y)t^{d_{\lambda
}},\qquad \lambda \vdash d_{\lambda }.
\end{equation}%
Using the bijection between traces of irreducible representations and
irreducible characters, it follows that there is an isomorphism of $\left(
GL(U)\times GL(V)\right) $-representations:
\begin{equation}
S^{k}(U\otimes V)\,\cong \,\bigoplus_{\lambda \vdash k,ht(\lambda )\leq
m}\,S_{\lambda }(U)\otimes S_{\lambda }(V).  \label{iso-1}
\end{equation}

A particular consequence of the isomorphism (\ref{iso-1}) is that if $%
\mathcal{G}\times G$ is a subgroup of $GL(U)\times GL(V)$, then the vector
space $\left( S^{k}(U\otimes V)\right) ^{\mathcal{G}\times G}$ of $\left(
\mathcal{G}\times G\right) $-invariants in $S^{k}(U\otimes V)$ enjoys the
following decomposition:
\begin{equation}
\left( S^{k}(U\otimes V)\right) ^{\mathcal{G}\times G}\;\cong
\;\bigoplus_{\lambda \vdash k,ht(\lambda )\leq m}\,(S_{\lambda }(U)^{%
\mathcal{G}})\otimes (S_{\lambda }(V)^{G}),  \label{decomposition}
\end{equation}%
since the action of $\mathcal{G}\times G$ on $S_{\lambda }(U)\otimes
S_{\lambda }(V)$ preserves the factors. Thus, \textit{in order to compute
the }$\left( \mathcal{G}\times G\right) $\textit{-invariants, one can
compute the} $\mathcal{G}$\textit{-invariants on all} $S_{\lambda }(U)$
\textit{and the }$G$\textit{-invariants on all }$S_{\lambda }(V)$\textit{,
and then combine the results}.

Given a partition $\lambda =(p_{1},\ldots ,p_{N})$, we define an integer $%
k\in \mathbb{Z}_{\geq 0}$ and a partition $\mu $ with $ht(\mu )\leq N-1$ by
\begin{equation}
\lambda \,=\,(k,\ldots ,k)\,+\,(k_{1},\ldots ,k_{N-1},0)\,:=\,(k^{N})+\mu .
\label{def-1}
\end{equation}%
Then, the restriction of $S_{\lambda }(V)$ to $SL(V)$ is isomorphic to $%
S_{\mu }(V)$, since $(k^{N})$ is the $k$-th tensor product of the
determinant representation. \textit{\c{C}a va sans dire}, if $ht(\lambda )>n$%
, then the definition of $S_{\lambda }(V)$ shows that it is the $0$%
-dimensional vector space.


\subsection{\label{p-bH}Application to $p$-centered Black Holes}

As in Sec.\ \ref{Intro}, let $G_{4}$ be the $U$-duality group acting on the
representation $R$ in which the (fluxes of the) Abelian $2$-form field
strengths and their duals sit, in the background of a $p$-centered black
hole solution in the corresponding $D=4$ Maxwell-Einstein (super)gravity
theory. Since the \textit{\textquotedblleft horizontal"} \cite{FMOSY-1}
group $SL_{h}(p)\equiv SL_{h}(p,\mathbb{R})$ acts on the labels of the
centers, in order to determine the invariants associated to the $p$-centered
BH one has to compute the invariants of $\mathcal{G}\times G=SL_{h}(p)\times
G_{4}$ on $U\otimes V=\mathbb{R}^{p}\otimes R=:\left( \mathbf{p},R\right) $.

The representation $S_{\lambda }(U)$, where $U=\mathbb{R}^{p}$, of $\mathcal{%
G}=SL_{h}(p)$, is irreducible (if non-zero), and there are very few cases in
which it is the trivial $1$-dimensional representation. In fact, recall that
$S_{\lambda }(V)=0$ if $ht(\lambda )>p$, whereas if $ht(\lambda )<p$ then $%
S_{\lambda }(V)$ is an irreducible representation of $GL(V)$, and hence also
of $SL(V)$. Thus $S_{\lambda }(V)^{\mathcal{G}}=0$, unless $S_{\lambda }(V)$
is a power of the $1$-dimensional determinant representation of $GL(V)$;
namely, unless the partition reads $\lambda =(a,\ldots ,a)=:(a^{p})$, in
which case one has
\begin{equation}
U=\mathbb{R}^{p}~\text{of~}\mathcal{G}=SL_{h}(p):\dim \left( S_{\lambda
}(U)^{SL_{h}(U)}\right) \,=\,1\quad \Longleftrightarrow \quad \lambda
\,=\,(a^{p})\,=\,\underbrace{(a,\ldots ,a)}_{p}
\end{equation}%
for some $a\in \mathbb{Z}_{\geq 0}$, and $\dim \left( S_{\lambda
}(U)^{SL_{h}(U)}\right) =0$ otherwise.

In virtue of formula (\ref{decomposition}), this implies that the invariants
of $SL_{h}(p)\times G_{4}$ in $S^{k}(\mathbb{R}^{p}\otimes R)$ must come
from the invariants of $G_{4}$ in $S_{\lambda }(R)$ where $\lambda =(a^{p})$%
. As $(a^{p})\vdash pa$, it thus also follows\footnote{%
In Sec.\ \ref{grass} we will discuss in some detail the $G_{4}$%
-representation $S_{(a^{p})}(R)$ which gives rise to all invariants.}
\begin{equation}
\dim \left( S^{pa}(\mathbb{R}^{p}\otimes R)^{SL_{h}(p)\times G_{4}}\right)
\,=\,\dim \left( S_{(a^{p})}(R)^{G_{4}}\right) ,  \label{res-1}
\end{equation}%
and \textit{there are no invariants}\footnote{%
Besides the above reasoning, another proof of this fact is the following one
: the group $SL(p,\mathbb{C})$ contains the matrices $\lambda I$ where $%
\lambda =e^{2\pi i/p}$, and the element $(\lambda I,I)\in SL(p)\times G_{4}$
acts as multiplication by the scalar $\lambda $ on $\mathbb{C}^{p}\otimes R$%
, and hence by $\lambda ^{k}$ on $S^{k}(\mathbb{C}^{p}\otimes R)$.}\textit{\
in }$S^{k}(\mathbb{R}^{p}\otimes R)$\textit{\ if }$k$\textit{\ is not a
multiple of }$p$\textit{.} So, if one has a degree-$k$ $\left(
SL_{h}(p)\times G_{4}\right) $-invariant homogeneous polynomial in the
representation $\mathbb{R}^{p}\otimes R$, then $k$ is a multiple of $p$ (the
converse surely does not hold; see \textit{e.g.} Tables (\ref{TableE7}), (%
\ref{TableSp6}), (\ref{TableSO(12)}), (\ref{TableSU(6)}) and (\ref{tab:stu})
below).\medskip\ 

Before explicitly analyzing some cases relevant to supergravity, let us
consider the lowest degrees of homogeneity : $k=2$ and $k=3$.

\subsubsection{\label{k=2}Homogeneity $k=2$}

In the case $k=2$, the partitions $\lambda $ with $\lambda \vdash 2$ are $%
\lambda =(2,0)=:2$ and $\lambda =(1,1)=:1^{2}$. Since $S_{2}\left( V\right)
=S^{2}V$ \ and $S_{1^{2}}=\wedge ^{2}V$, one obtains (provided $ht\left(
\lambda \right) \leq 2\leq \min \left( \dim U,\dim V\right) $):%
\begin{equation}
S^{2}\left( U\otimes V\right) \cong \left( S^{2}U\right) \otimes \left(
S^{2}V\right) \oplus \left( \wedge ^{2}U\right) \otimes \left( \wedge
^{2}V\right) .  \label{decomp-1}
\end{equation}

A particular case, in which the term $\left( S^{2}U\right) \otimes \left(
S^{2}V\right) $ does not yield any invariant, is provided by $2$-centered ($%
p=2$) BHs in the framework under consideration, namely for $p=2$: $U\otimes
V=\mathbb{R}^{2}\otimes R$ of $\mathcal{G}\times G=SL_{h}(2,\mathbb{R}%
)\times G_{4}$. As both $SL_{h}(2,\mathbb{R})$ and $G_{4}$ have an invariant
in $\wedge ^{2}\mathbb{R}^{2}\cong \mathbb{R}$ and in $\wedge ^{2}R$,
respectively (namely, both the fundamental spin $s=1/2$ irrep. $\mathbf{2}$
of $SL_{h}(2,\mathbb{R})$ and the irrep. $R$ of $G_{4}$ are \textit{%
symplectic}) one obtains one invariant from the term $\left( \wedge
^{2}U\right) \otimes \left( \wedge ^{2}V\right) $ of (\ref{decomp-1}), given
by the \textit{symplectic product} $\mathcal{W}$ in $R$ of $G_{4}$, namely
\cite{FMOSY-1,ADFMT-1,FMY-CV-1} ($a,b=1,2$, $M,N=1,...,\dim R$):%
\begin{equation}
\mathcal{W}:=\left( Q_{1},Q_{2}\right) :=\mathbb{C}_{MN}Q_{1}^{M}Q_{2}^{N}=%
\frac{1}{2}\epsilon ^{ab}\mathbb{C}_{MN}Q_{a}^{M}Q_{b}^{N},  \label{W}
\end{equation}%
where $\epsilon ^{ab}$ is the Ricci-Levi-Civita symbol of $SL_{h}\left( 2,%
\mathbb{R}\right) $. When $\mathcal{W}\neq 0$, the charge vectors $Q_{1}$
and $Q_{2}$ (respectively pertaining to BH centers $1$ and $2$) are \textit{%
mutually non-local}, and the distance between the two centers in the BPS $2$%
-centered system is fixed \cite{Denef}. No other algebraically independent
invariant polynomial homogeneous of degree $k=2$ arise, since the
representations $U=\mathbb{R}^{2}=:\mathbf{2}$ of $SL_{h}(2,\mathbb{R})$ and
$V=R$ of $G_{4}$ are irreducible, and thus there are no other invariants in $%
S^{2}U$ and in $S^{2}V$.

As discussed at the end of Sec.\ 3 of \cite{ADFMT-1}, some $SL_{h}\left( p,%
\mathbb{R}\right) $-covariant structures for $p\geqslant 3$ can be directly
inferred from the $2$-centered ones. Indeed, the $2$-centered representation
of spin $s=J/2$ of $SL_{h}\left( 2,\mathbb{R}\right) $ is in general
replaced by the completely symmetric rank-$J$ tensor representation\footnote{%
In the case of $GL(p,\mathbb{R})$, this is given by $S_{\lambda }(V)$ (\ref%
{S_k}) with $V=\mathbb{R}^{p}=:\mathbf{p}$ and $\lambda :=J:=(J,\underbrace{%
0,\ldots ,0}_{p-1})$; see below (\ref{dim-S-lambda}). The same holds for $%
SL(p,\mathbb{R})$.} $S^{J}\mathbf{p}$ of $SL_{h}\left( p,\mathbb{R}\right) $%
. On the other hand, for $p$ centers $\mathcal{W}$ (\ref{W}) generally sits
in the $\left( \wedge ^{2}\mathbf{p},\mathbf{1}\right) $ of $SL_{h}\left( p,%
\mathbb{R}\right) \times G_{4}$, where $\wedge ^{2}\mathbf{p}$ is the rank-$%
2 $ antisymmetric tensor representation\footnote{%
In the case of $GL(p,\mathbb{R})$, this is given by $S_{\lambda }(V)$ (\ref%
{S_1^h}) with $V=\mathbb{R}^{p}=:\mathbf{p}$ and $\lambda :=1^{2}$; see
below (\ref{dim-S-lambda}). The same holds for $SL(p,\mathbb{R})$.} (which,
in the case $p=2$, becomes a singlet).

\subsubsection{\label{k=3}Homogeneity $k=3$ for $G_{4}=E_{7}$}

In the case $k=3$, the partitions $\lambda $ with $\lambda \vdash 3$ are $%
\lambda =(3,0,0)=:3$, $\lambda =(2,1,0)=:(2,1)$ and $\lambda =(1,1,1)=:1^{3}$%
. Since $S_{3}\left( V\right) =S^{3}V$ \ and $S_{1^{3}}=\wedge ^{3}V$, the $%
GL(V)$-representation $S_{(2,1)}\left( V\right) $ is obtained by the
decomposition\footnote{%
This is generalized to $V^{\otimes n}$ (for a generic $n$) \textit{e.g.} in
\cite{Procesi}, 9.3.1.} (\textit{cfr. e.g.} \cite{Procesi}, 9.3.1)%
\begin{equation}
V^{\otimes 3}:=V\otimes V\otimes V\cong \left( S^{3}V\right) \oplus \left(
S_{(2,1)}\left( V\right) \right) ^{\oplus 2}\oplus \wedge ^{3}V.  \label{V3}
\end{equation}

The simplest example is provided once again by $V=\mathbb{R}^{2}=:\mathbf{2}$
of $SL_{h}(2,\mathbb{R})$, for which it holds%
\begin{equation}
\mathbf{2}\otimes \mathbf{2}\otimes \boldsymbol{2}\cong \left( \mathbf{3}%
\oplus \mathbf{1}\right) \otimes \boldsymbol{2}\cong \left( \mathbf{4}\oplus
\mathbf{2}\right) \oplus \boldsymbol{2},
\end{equation}%
where $\mathbf{4}=:S^{3}V$ is the spin $s=3/2$ of $SL_{h}(2,\mathbb{R})$
itself, consistent with the Clebsch-Gordan formula for this group.

Another example, in which we also exploit the physicists' notation of
representations by means of their dimension, is provided by $V=V\left(
\lambda _{7}\right) =:\mathbf{56}$ (fundamental) irrep. of $G_{4}=E_{7}$. In
this case, the following decomposition holds\footnote{%
The weights/roots standard notation of irreps. is used throughout.}:%
\begin{eqnarray}
\underset{\left( \mathbf{56}\otimes \mathbf{56}\otimes \mathbf{56}\right)
_{s}}{S^{3}V\left( \lambda _{7}\right) } &\cong &\underset{\mathbf{24320}}{%
V\left( 3\lambda _{7}\right) }\oplus \underset{\mathbf{6480}}{V\left(
\lambda _{1}+\lambda _{7}\right) }\oplus \underset{\mathbf{56}}{V\left(
\lambda _{7}\right) };  \label{1} \\
\underset{\left( \mathbf{56}\otimes \mathbf{56}\otimes \mathbf{56}\right)
_{a}}{\wedge ^{3}V\left( \lambda _{7}\right) } &\cong &\underset{\mathbf{%
27664}}{V\left( \lambda _{5}\right) }\oplus \underset{\mathbf{56}}{V\left(
\lambda _{7}\right) }.  \label{2}
\end{eqnarray}%
On the other hand:%
\begin{equation}
\underset{\mathbf{56}\otimes \mathbf{56}}{V^{\otimes 2}}:=V\otimes V\cong
\underset{\left( \mathbf{56}\otimes \mathbf{56}\right) _{s}}{\left(
S^{2}V\right) }\oplus \underset{\left( \mathbf{56}\otimes \mathbf{56}\right)
_{a}}{\wedge ^{2}V}=\underset{\mathbf{1463}\oplus \mathbf{133}}{\left(
V\left( 2\lambda _{7}\right) \oplus V\left( \lambda _{1}\right) \right) }%
\oplus \underset{\mathbf{1539}\oplus \mathbf{1}}{\left( V\left( \lambda
_{6}\right) \oplus V\left( 0\right) \right) }.  \label{3}
\end{equation}%
Thus, tensoring once more with $V\left( \lambda _{7}\right) $, one obtains%
\begin{eqnarray}
\underset{\mathbf{1463}\otimes \mathbf{56}}{V\left( 2\lambda _{7}\right)
\otimes V\left( \lambda _{7}\right) ~} &\cong &\underset{\mathbf{24320}}{%
~V\left( 3\lambda _{7}\right) }\oplus \underset{\mathbf{51072}}{V\left(
\lambda _{6}+\lambda _{7}\right) }\oplus \underset{\mathbf{6480}}{V\left(
\lambda _{1}+\lambda _{7}\right) }\oplus \underset{\mathbf{56}}{V\left(
\lambda _{7}\right) };  \label{4} \\
\underset{\mathbf{1539}\otimes \mathbf{56}}{V\left( \lambda _{6}\right)
\otimes V\left( \lambda _{7}\right) ~} &\cong &\underset{\mathbf{51072}}{%
~V\left( \lambda _{6}+\lambda _{7}\right) }\oplus \underset{\mathbf{27664}}{%
V\left( \lambda _{5}\right) }\oplus \underset{\mathbf{6480}}{V\left( \lambda
_{1}+\lambda _{7}\right) }\oplus \underset{\mathbf{912}}{V\left( \lambda
_{2}\right) }\oplus \underset{\mathbf{56}}{V\left( \lambda _{7}\right) };
\label{5} \\
\underset{\mathbf{133}\otimes \mathbf{56}}{V\left( \lambda _{1}\right)
\otimes V\left( \lambda _{7}\right) ~} &\cong &\underset{\mathbf{6480}}{%
~V\left( \lambda _{1}+\lambda _{7}\right) }\oplus \underset{\mathbf{512}}{%
V\left( \lambda _{2}\right) }\oplus \underset{\mathbf{56}}{V\left( \lambda
_{7}\right) }.  \label{6}
\end{eqnarray}%
Thus, (\ref{V3}) and (\ref{1})-(\ref{6}) yield%
\begin{equation}
S_{(2,1)}\left( V\left( \lambda _{7}\right) \right) \cong \underset{\mathbf{%
51072}}{~V\left( \lambda _{6}+\lambda _{7}\right) }\oplus \underset{\mathbf{%
6480}}{V\left( \lambda _{1}+\lambda _{7}\right) }\oplus \underset{\mathbf{912%
}}{V\left( \lambda _{2}\right) }\oplus \underset{\mathbf{56}}{V\left(
\lambda _{7}\right) }.
\end{equation}

Therefore, we obtained that%
\begin{equation}
\underset{\mathbf{1}}{V(0)}\notin \left\{
\begin{array}{l}
S^{3}V\left( \lambda _{7}\right) ; \\
S_{(2,1)}\left( V\left( \lambda _{7}\right) \right) ; \\
\wedge ^{3}V\left( \lambda _{7}\right) ,%
\end{array}%
\right.
\end{equation}%
and thus there are no $E_{7}$-invariants in $S_{\lambda }\left( V\left(
\lambda _{7}\right) \right) $ if $\lambda \vdash 3$. More in general,
\textit{there are no} $E_{7}$\textit{-invariants on} $V\left( \lambda
_{7}\right) ^{\otimes n}$ \textit{for }$n$\textit{\ odd}. Since $S_{\lambda
}\left( V\right) \subset V^{\otimes n}$ when $\lambda \vdash n$, it follows
that there are no $E_{7}$-invariants in $S_{\lambda }\left( V\left( \lambda
_{7}\right) \right) $ when $\lambda $ is a partition of an \textit{odd}
(positive) integer $n$.

In other words, \textit{there are no invariant polynomials in the
fundamental representation} $V\left( \lambda _{7}\right) =:\mathbf{56}$
\textit{of} $E_{7}$ \textit{with an} \textit{odd} \textit{homogeneity degree}%
, as also confirmed by the treatment of Sec.\ \ref{nBHe7}; more in general,
this will hold \textit{at least} for all the (simple and semi-simple) groups
\textquotedblleft of type $E_{7}$" which we will consider : \textit{there
are no invariant polynomials in the relevant irrep. }$R$ \textit{of} $G_{4}$
\textit{with an odd homogeneity degree}\footnote{%
The reason can be traced back to the fact that $-I$ on $R$ belongs to $G_{4}$%
. For instance, it can be checked that the $-I$ in the $\mathbf{56}$ of $%
E_{7}$ preserves the symplectic metric $\mathbb{C}_{\left[ MN\right] }$ in $%
\mathbf{56}_{a}^{\otimes 2}$ and the quartic symmetric tensor $K_{\left(
MNPQ\right) }$ in $\mathbf{56}_{s}^{\otimes 4}$ ($M,N,P,Q=1,..,56$).}.

\subsection{\label{Examples}Examples}

We now consider explicit examples, relevant for $p$-centered ($p\geqslant 2$%
) black holes in some $D=4$ Maxwell-Einstein (super)gravity theories, with
generalized electric-magnetic ($U$-)duality group $G_{4}$; as done above, we
denote the relevant $G_{4}$-representation in which the (fluxes of the)
Abelian $2$-form field strengths (and their duals) sit by\footnote{%
It is worth pointing out that the irrep. $R$ is real for the very
non-compact real forms of $G_{4}$ pertaining to the relevant $U$-duality
groups, while usually for the other (non-compact) real forms it is
pseudo-real (quaternionic). This reality property can \textit{e.g.} be
inferred from the corresponding (symmetric) embeddings into $G_{3}$, the
relevant $U$-duality symmetry in $D=3$ space-time dimensions.
\par
As an example, let us consider the fundamental representation $R=\mathbf{56}$
of $E_{7}$ : it is real for the relevant non-compact real forms $E_{7(7)}$ (%
\textit{split}) and $E_{7(-25)}$ (minimally non-compact), while it is
pseudo-real (quaternionic) for $E_{7(-133)}$ and $E_{7(-5)}$. Indeed, while $%
E_{7(7)}$ and $E_{7(-25)}$ respectively embed into $E_{8\left( 8\right) }$
and $E_{8(-24)}$ through a $SL(2,\mathbb{R})$ commuting factor:%
\begin{equation*}
E_{8(8)}\supset E_{7(7)}\times SL(2,\mathbb{R}),~~E_{8(-24)}\supset
E_{7(-25)}\times SL(2,\mathbb{R}),
\end{equation*}%
$E_{7(-133)}$ and $E_{7(-5)}$ embed into $E_{8\left( -24\right) }$ and $%
E_{8(8)}$ through an $SU(2)$ factor:%
\begin{eqnarray*}
E_{8(-24)} &\supset &E_{7(-133)}\times SU(2),~~E_{8(-24)}\supset
E_{7(-5)}\times SU(2); \\
E_{8(8)} &\supset &E_{7(-5)}\times SU(2).
\end{eqnarray*}%
} $V=R$, and we will specify it case by case.

\begin{table}[t]
\begin{center}
\begin{tabular}{|c||c|c|c|}
\hline
$%
\begin{array}{c}
\\
J_{3}%
\end{array}%
$ & $%
\begin{array}{c}
\\
G_{4} \\
~~%
\end{array}%
$ & $%
\begin{array}{c}
\\
R \\
~~%
\end{array}%
$ & $%
\begin{array}{c}
\\
\mathcal{N} \\
~~%
\end{array}%
$ \\ \hline\hline
$%
\begin{array}{c}
\\
J_{3}^{\mathbb{O}},~J_{3}^{\mathbb{O}_{s}} \\
~%
\end{array}%
$ & $E_{7\left( -25\right) },~E_{7(7)}~$ & $\mathbf{56}$ & $2,~8~$ \\ \hline
$%
\begin{array}{c}
\\
J_{3}^{\mathbb{H}},~J_{3}^{\mathbb{H}_{s}} \\
~%
\end{array}%
$ & $SO^{\ast }(12),~SO(6,6)$ & $\mathbf{32}^{(\prime )}$ & $2~$or$~6,~0$ \\
\hline
$%
\begin{array}{c}
\\
J_{3}^{\mathbb{C}},~J_{3}^{\mathbb{C}_{s}},~M_{1,2}\left( \mathbb{O}\right)
\\
~%
\end{array}%
$ & $SU\left( 3,3\right) ,~SL(6,\mathbb{R}),~SU(1,5)$ & $\mathbf{20}$ & $%
2,~0,~5$ \\ \hline
$%
\begin{array}{c}
\\
J_{3}^{\mathbb{R}} \\
~%
\end{array}%
$ & $Sp\left( 6,\mathbb{R}\right) $ & $\mathbf{14}^{\prime }$ & $2~$ \\
\hline
$%
\begin{array}{c}
\\
\mathbb{R} \\
(t^{3}\text{~model})~~%
\end{array}%
$ & $SL\left( 2,\mathbb{R}\right) $ & $\mathbf{4}$ & $2$ \\ \hline
$%
\begin{array}{c}
\\
R\oplus \mathbf{\Gamma }_{m-1,n-1} \\
~%
\end{array}%
$ & $SL\left( 2,\mathbb{R}\right) \times SO(m,n)$ & $\left( \mathbf{2,m+n}%
\right) $ & $%
\begin{array}{c}
2~\left( m~\text{or~}n=2\right) \\
4~(m~\text{or~}n=6) \\
~0~\text{otherwise}%
\end{array}%
$ \\ \hline
\end{tabular}%
\end{center}
\caption{Simple and semi-simple, \textit{non-degenerate} $U$-duality groups $%
G_{4}$ \textquotedblleft of type $E_{7}$" \protect\cite{brown}. The relevant
symplectic irrep. $R$ of $G_{4}$ is also reported. Note that the $G_{4}$
related to split composition algebras $\mathbb{O}_{s}$, $\mathbb{H}_{s}$, $%
\mathbb{C}_{s}$ is the \textit{maximally non-compact} (\textit{split}) real
form of the corresponding compact Lie group. The corresponding scalar
manifolds are the \textit{symmetric} spaces $\frac{G_{4}}{H_{4}}$, where $%
H_{4}$ is the maximal compact subgroup (with symmetric embedding) of $G_{4}$%
. The number of supercharges $\mathcal{N}$ of the resulting supergravity
theory in $D=4$ is also listed. The $D=5$ uplift of the $t^{3}$ model (based
on $J_{3}=\mathbb{R}$) is the \textit{pure} $\mathcal{N}=2$, $D=5$
supergravity. $J_{3}^{\mathbb{H}}$ is related to both $8$ and $24$
supersymmetries, because the corresponding supergravity theories share the
very same bosonic sector \protect\cite{GST,twin,Gnecchi-1}. }
\end{table}

In particular, we here consider the class of groups \textquotedblleft of
type $E_{7}$" \cite{brown} which can be characterized as conformal groups of
rank-$3$, \textit{simple} Euclidean Jordan algebras $J_{3}^{\mathbb{A}}$ or $%
J_{3}^{\mathbb{A}_{s}}$, or equivalently as the automorphism group of the
Freudenthal triple system (FTS) $\mathfrak{M}\left( J_{3}\right) $
constructed over such algebras \cite{conf}:%
\begin{equation}
G_{4}=Conf\left( J_{3}\right) =Aut\left( \mathfrak{M}\left( J_{3}\right)
\right) .
\end{equation}%
$\mathbb{A}$ denotes the division algebras $\mathbb{A}=\mathbb{O},\mathbb{H},%
\mathbb{C},\mathbb{R}$, while $\mathbb{A}_{s}$ denotes the corresponding
split composition algebras $\mathbb{A}_{s}=\mathbb{O}_{s},\mathbb{H}_{s},%
\mathbb{C}_{s},\mathbb{R}$. The representation $R$ pertains to $\mathfrak{M}%
\left( J_{3}\right) $, and its dimension is $6q+8$, where the parameter $%
q=\dim _{\mathbb{R}}\mathbb{A}_{(s)}=8,4,2,1$ for $\mathbb{A}_{(s)}=\mathbb{O%
}_{(s)},\mathbb{H}_{(s)},\mathbb{C}_{(s)},\mathbb{R}$, respectively. These
class of groups \textquotedblleft of type $E_{7}$" has been recently studied
as $U$-duality symmetries in the context of $D=4$ locally supersymmetric
theories of gravity in \cite{Duff-E7,FK-E7,FKM-E7}, as well as gauge (and
global) symmetries in particular $D=3$ gauge theories \cite{FGT}.

An exception is provided by the $stu$ model \cite{STU} (Sec.\ \ref{STU-gen}%
), whose \textit{triality symmetry} is exploited within a particular case in
Sec.\ \ref{stu}.

From Sec.\ \ref{p-bH}, it is here worth recalling that in general \textit{%
there are no polynomial invariants of }$\left( \mathbf{p},R\right) $ of $%
SL_{h}\left( p,\mathbb{R}\right) \times G_{4}$\textit{\ with homogeneity
degree }$k$\textit{\ if }$k$\textit{\ is not a multiple of }$p$\textit{.}

\subsubsection{\label{nBHe7}{$G_{4}=E_{7}$, }$R=\mathbf{56}$}

This is the prototypical case of groups \textquotedblleft of type $E_{7}$"
\cite{brown}. In supergravity, this is related to the $D=4$ theories with
symmetric scalar manifold, based on the FTS $\mathfrak{M}\left( J_{3}^{%
\mathbb{O}}\right) $ (exceptional $\mathcal{N}=2$ Maxwell-Einstein theory,
with $G_{4}=E_{7(-25)}$ \cite{GST}) and $\mathfrak{M}\left( J_{3}^{\mathbb{O}%
_{s}}\right) $ ($\mathcal{N}=8$ maximal supergravity, with $G_{4}=E_{7(7)}$
\cite{CJ-1,N=8}), where $J_{3}^{\mathbb{O}}$ and $J_{3}^{\mathbb{O}_{s}}$
are rank-$3$ Euclidean Jordan algebras over the octonions $\mathbb{O}$ and
split octonions $\mathbb{O}_{s}$, respectively.

The dimension $\dim S_{(a^{p})}(R)^{E_{7}}$ for the partition $\lambda
=a^{p} $ and $R=V\left( \lambda _{7}\right) =:\mathbf{56}$ (fundamental
irrep.) can be computed \textit{e.g.} by using the software \texttt{LiE}
\cite{LiE}, typing the command\footnote{%
In \texttt{LiE}, one first increases the maximal size by typing the command
\textquotedblleft \texttt{maxobjects 99999999}".}
\begin{equation}
\mbox{plethysm([$a$,\ldots,$a$],[0,0,0,0,0,0,1],E7)[1]}.  \label{nBHE7-input}
\end{equation}%
The \textquotedblleft $\lbrack 1]$" at the end corresponds to the lowest
representation. The output of the command is an integer, which we denote by $%
d$, times $X[b_{1},\ldots ,b_{7}]$, where $X[b_{1},\ldots ,b_{7}]$ indicates
the representation with highest weight $b_{1}\lambda _{1}+\cdots
+b_{7}\lambda _{7}$, the $\lambda _{i}$ being the fundamental weights ($%
i=1,...,7$). If all $b_{i}$'s are zero, then one has found polynomial
invariants of homogeneity degree $pa$ in $p\dim R=56p$ variables; the real
dimension of the vector space of such invariants is given by (recall (\ref%
{res-1}))%
\begin{equation}
\dim \left[ S_{\lambda =a^{p}}\left( V\left( \lambda _{7}\right) \right) %
\right] ^{E_{7}}=\dim \left[ S^{pa}\left( p,V\left( \lambda _{7}\right)
\right) \right] ^{SL_{h}\left( p,\mathbb{R}\right) \times E_{7}}=\dim \left[
S^{pa}\left( \mathbf{p},\mathbf{56}\right) \right] ^{SL_{h}\left( p,\mathbb{R%
}\right) \times E_{7}}=:d.
\end{equation}

By perusing the first few $a$'s for the first few $p$'s, one gets the
following table\footnote{%
The result $\dim \left[ S_{\lambda =0^{p}}\left( V\right) \right] ^{G_{4}}=1$
always trivially refers to a numerical constant.}:
\begin{equation}
\begin{array}{|c|c|c|c|c|c|c|c|c|c|c|c|c|}
\hline
E_{7},~\mathbf{56} & a= & 0 & 1 & 2 & 3 & 4 & 5 & 6 & 7 & 8 & 9 & 10 \\
\hline
p=2 & d= & 1 & 1 & 1 & 2 & 3 & 3 & 5 & 6 & 7 & 9 & 11 \\ \hline
p=3 & d= & 1 & 0 & 0 & 0 & 5 & 0 & 1 & 0 & 46 &  &  \\ \hline
p=4 & d= & 1 & 1 & 1 & 4 & 14 & 35 &  &  &  &  &  \\ \hline
p=5 & d= & 1 & 0 & 0 & 0 & 31 &  &  &  &  &  &  \\ \hline
p=6 & d= & 1 & 1 & 2 & 10 &  &  &  &  &  &  &  \\ \hline
p=7 & d= & 1 & 0 & 2 &  &  &  &  &  &  &  &  \\ \hline
p=8 & d= & 1 & 1 &  &  &  &  &  &  &  &  &  \\ \hline
\end{array}
\label{TableE7}
\end{equation}%
(throughout the treatment, the blank entries are seemingly not accessible
with the computing facilities available to us.)

In the $2$-centered case ($p=2$), $\dim S_{1^{2}}(\mathbf{56})^{E_{7}}=1$
corresponds to $\mathcal{W}$ (\ref{W}). The interpretation of the other
results is as follows:%
\begin{eqnarray}
&&%
\begin{array}{lll}
\dim S_{2^{2}}(\mathbf{56})^{E_{7}}=1 & : & \mathcal{W}^{2} \\
\dim S_{3^{2}}(\mathbf{56})^{E_{7}}=2 & : & \mathcal{W}^{3},~\mathbf{I}_{6}
\\
\dim S_{4^{2}}(\mathbf{56})^{E_{7}}=3 & : & \mathcal{W}^{4},~\mathbf{I}_{6}%
\mathcal{W},~\text{Tr}\left( \mathbf{I}^{2}\right) \\
\dim S_{5^{2}}(\mathbf{56})^{E_{7}}=3 & : & \mathcal{W}^{5},~\mathbf{I}_{6}%
\mathcal{W}^{2},~\text{Tr}\left( \mathbf{I}^{2}\right) \mathcal{W} \\
\dim S_{6^{2}}(\mathbf{56})^{E_{7}}=5 & : & \mathcal{W}^{6},~\mathbf{I}%
_{6}^{2},~\text{Tr}\left( \mathbf{I}^{2}\right) \mathcal{W}^{2},~\text{Tr}%
\left( \mathbf{I}^{3}\right) ,~\mathbf{I}_{6}\mathcal{W}^{3} \\
\dim S_{7^{2}}(\mathbf{56})^{E_{7}}=6 & : & \mathcal{W}^{7},~\mathbf{I}%
_{6}^{2}\mathcal{W},~\mathbf{I}_{6}\text{Tr}\left( \mathbf{I}^{2}\right) ,~%
\text{Tr}\left( \mathbf{I}^{3}\right) \mathcal{W},~\text{Tr}\left( \mathbf{I}%
^{2}\right) \mathcal{W}^{3},~\mathbf{I}_{6}\mathcal{W}^{4} \\
\dim S_{8^{2}}(\mathbf{56})^{E_{7}}=7 & : & \left\{
\begin{array}{l}
\mathcal{W}^{8},~\mathbf{I}_{6}^{2}\mathcal{W}^{2},~\mathbf{I}_{6}\text{Tr}%
\left( \mathbf{I}^{2}\right) \mathcal{W},~\text{Tr}\left( \mathbf{I}%
^{3}\right) \mathcal{W}^{2}, \\
\text{Tr}\left( \mathbf{I}^{2}\right) \mathcal{W}^{4},~\mathbf{I}_{6}%
\mathcal{W}^{5},~\text{Tr}^{2}\left( \mathbf{I}^{2}\right)%
\end{array}%
\right. \\
\dim S_{9^{2}}(\mathbf{56})^{E_{7}}=9 & : & \left\{
\begin{array}{l}
\mathcal{W}^{9},~\mathbf{I}_{6}^{2}\mathcal{W}^{3},~\mathbf{I}_{6}\text{Tr}%
\left( \mathbf{I}^{2}\right) \mathcal{W}^{2},~\text{Tr}\left( \mathbf{I}%
^{3}\right) \mathcal{W}^{3}, \\
\text{Tr}\left( \mathbf{I}^{2}\right) \mathcal{W}^{5},~\mathbf{I}_{6}%
\mathcal{W}^{6},~\text{Tr}^{2}\left( \mathbf{I}^{2}\right) \mathcal{W},~%
\text{Tr}\left( \mathbf{I}^{3}\right) \mathbf{I}_{6},~\mathbf{I}_{6}^{3}%
\end{array}%
\right. \\
\dim S_{10^{2}}(\mathbf{56})^{E_{7}}=11 & : & \left\{
\begin{array}{l}
\mathcal{W}^{10},~\mathbf{I}_{6}^{2}\mathcal{W}^{4},~\mathbf{I}_{6}\text{Tr}%
\left( \mathbf{I}^{2}\right) \mathcal{W}^{3},~\text{Tr}\left( \mathbf{I}%
^{3}\right) \mathcal{W}^{4},~\text{Tr}\left( \mathbf{I}^{2}\right) \mathcal{W%
}^{6}, \\
\mathbf{I}_{6}\mathcal{W}^{7},~\text{Tr}^{2}\left( \mathbf{I}^{2}\right)
\mathcal{W}^{2},~\text{Tr}\left( \mathbf{I}^{3}\right) \mathbf{I}_{6}%
\mathcal{W},~\mathbf{I}_{6}^{3}\mathcal{W},~\mathbf{I}_{6}^{2}\text{Tr}%
\left( \mathbf{I}^{2}\right) ,~\text{Tr}\left( \mathbf{I}^{2}\right) \text{Tr%
}\left( \mathbf{I}^{3}\right) ,%
\end{array}%
\right.%
\end{array}
\notag \\
&&  \label{TableE7-p=2}
\end{eqnarray}%
where the $2$-centered polynomial invariants\footnote{%
As discussed at the end of Sec.\ 3 of \cite{ADFMT-1}, for $p$ centers $%
\mathbf{I}_{6}$, as $\mathcal{W}$ (\ref{W}), generally sits in the $\left(
\wedge ^{2}\mathbf{p},\mathbf{1}\right) $ of $SL_{h}\left( p,\mathbb{R}%
\right) \times G_{4}$.} $\mathbf{I}_{6}$ (degree $6$), Tr$\left( \mathbf{I}%
^{2}\right) $ (degree $8$) and Tr$\left( \mathbf{I}^{3}\right) $ (degree $12$%
) have been firstly introduced in \cite{FMOSY-1}, and then studied in this
very case in \cite{ADFMT-1}. Note that no polynomial $2$-centered invariants
in the $\mathbf{56}$ of $E_{7}$ exist with an \textit{odd} homogeneity
degree, consistent with the observation made in Sec.\ \ref{k=3}. The
interpretation (\ref{TableE7-p=2}) of the $p=2$ row of Table (\ref{TableE7})
is an evidence for the fact that the set%
\begin{equation}
\left\{ \mathcal{W},\mathbf{I}_{6},\text{Tr}\left( \mathbf{I}^{2}\right) ,%
\text{Tr}\left( \mathbf{I}^{3}\right) \right\}  \label{set-1}
\end{equation}%
is a complete basis for the ring of polynomial invariants of $\left( \mathbf{%
2},\mathbf{56}\right) $ of $SL_{h}\left( 2,\mathbb{R}\right) \times E_{7}$,
and it is \textit{finitely generating}, namely all higher order polynomial
invariants are simply polynomials in the polynomials of the set (\ref{set-1}%
) itself \cite{Kac}.

In the $3$-centered case ($p=3$), Table (\ref{TableE7}) yields that there
are no $E_{7}$-invariants for the partitions $\lambda =1^{3}$, $2^{3}$, $%
3^{3}$ and hence there are no polynomial invariants of $\left( \mathbf{3},%
\mathbf{56}\right) $ of $SL_{h}\left( 3,\mathbb{R}\right) \times E_{7}$ with
homogeneity degree $\leq 10$. The lowest possible degree is $12$, at which
\texttt{Lie} finds $5$ invariants. The absence of an invariant corresponding
to the partition $\lambda =1^{3}$, \textit{i.e.} of a \textquotedblleft $3$%
-centered analogue" of $\mathcal{W}$ (\ref{W}) can be explained by the fact
that $\mathbf{1}\notin \mathbf{56}_{a}^{\otimes 3}$ (as mentioned, no
invariant polynomials in the $\mathbf{56}$ of $E_{7}$ with an \textit{odd}
homogeneity degree exist at all). Then, one invariant of degree $18$, and as
many as $46$ invariants of degree $24$, are found.

In the $4$-centered case ($p=4$), there is an $E_{7}$-invariant of degree $4$
(the lowest possible degree). It can be regarded as the \textquotedblleft $4$%
-centered analogue" of $\mathcal{W}$ (\ref{W}), whose existence can be
explained by the fact that $\exists !\mathbf{1}\in \mathbf{56}_{a}^{\otimes
4}$, given by the complete antisymmetrization of the product of two
symplectic metrics $\mathbb{C}_{MN}$ of $\mathbf{56}$, such that ($a=1,...,4$%
, $M=1,...,56$)%
\begin{equation}
\dim S_{1^{4}}(\mathbf{56})^{E_{7}}=1:\mathcal{W}_{p=4}:=\frac{1}{4!}\mathbb{%
C}_{[MN}\mathbb{C}_{PQ]}\epsilon ^{abcd}Q_{a}^{M}Q_{b}^{N}Q_{c}^{P}Q_{d}^{Q}.
\label{W-p=4}
\end{equation}%
Thus, $\mathcal{W}_{p=4}$ (\ref{W-p=4}) is the unique polynomial invariant
of $\left( \mathbf{4},\mathbf{56}\right) $ of $SL_{h}\left( 4,\mathbb{R}%
\right) \otimes E_{7}$ with homogeneity degree $4$. Its square yields the
unique polynomial invariant of $\left( \mathbf{4},\mathbf{56}\right) $ of $%
SL_{h}\left( 4,\mathbb{R}\right) \times E_{7}$ with homogeneity degree $8$,
as given by Table (\ref{TableE7}): $\dim S_{2^{4}}(\mathbf{56})^{E_{7}}=1$.

In the $5$-centered case ($p=5$), there are no invariants of degree $\leq 15$%
, since the partitions $\lambda =1^{5}$, $2^{5}$ and $3^{5}$ do not yield
any invariant for $E_{7}$. Once again, the absence of an invariant
corresponding to the partition $\lambda =1^{5}$, \textit{i.e.} of a
\textquotedblleft $5$-centered analogue" of $\mathcal{W}$ (\ref{W}), can be
explained by the fact that $\mathbf{1}\notin \mathbf{56}_{a}^{\otimes 5}$.

Finally, for the $p=6$ and $8$ -centered cases, we see that there is a
unique polynomial invariant of $\left( \mathbf{p},\mathbf{56}\right) $ of $%
SL_{h}\left( p,\mathbb{R}\right) \times E_{7}$ (corresponding to the
partition $\lambda =1^{p}$); again, for $p=6$ and $8$ it can be regarded as
the \textquotedblleft $p$-centered analogue" of $\mathcal{W}$ (\ref{W}),
whose existence can be explained by the fact that $\exists !\mathbf{1}\in
\mathbf{56}_{a}^{\otimes 6}$ and $\exists !\mathbf{1}\in \mathbf{56}%
_{a}^{\otimes 8}$, given by the complete antisymmetrization of the product
of $p=6,8$ symplectic metrics $\mathbb{C}_{MN}$ of $\mathbf{56}$, such that%
\begin{eqnarray}
\dim S_{1^{6}}(\mathbf{56})^{E_{7}} &=&1:\mathcal{W}_{p=6}:=\frac{1}{6!}%
\mathbb{C}_{[MN}\mathbb{C}_{PQ}\mathbb{C}_{RS]}\epsilon
^{abcdef}Q_{a}^{M}Q_{b}^{N}Q_{c}^{P}Q_{d}^{Q}Q_{e}^{R}Q_{f}^{S};  \label{uno}
\\
\dim S_{1^{8}}(\mathbf{56})^{E_{7}} &=&1:\mathcal{W}_{p=8}:=\frac{1}{8!}%
\mathbb{C}_{[MN}\mathbb{C}_{PQ}\mathbb{C}_{RS}\mathbb{C}_{TU]}\epsilon
^{abcdefgh}Q_{a}^{M}Q_{b}^{N}Q_{c}^{P}Q_{d}^{Q}Q_{e}^{R}Q_{f}^{S}Q_{g}^{T}Q_{h}^{U},
\notag  \label{due} \\
&&
\end{eqnarray}%
where the \textit{\textquotedblleft horizontal"} $a$-indices range over $%
1,...,6$ and $1,...,8$ in (\ref{uno}) and (\ref{due}), respectively.


\subsubsection{{$G_{4}=Sp(6,\mathbb{R})$, }$R=\mathbf{14}^{\prime }$}

In supergravity, this is related to the $D=4$ theory with symmetric scalar
manifold, based on the FTS $\mathfrak{M}\left( J_{3}^{\mathbb{R}}\right) $,
namely the magic $\mathcal{N}=2$ Maxwell-Einstein theory over $J_{3}^{%
\mathbb{R}}$ (the rank-$3$ Euclidean Jordan algebras over the reals $\mathbb{%
R}$ \cite{GST}).

In this case, the relevant $Sp(6,\mathbb{R})$-representation is\footnote{%
There are actually two irreducible representations of $Sp(6,\mathbb{R})$
with dimension $14$ : the rank-$2$ antisymmetric skew-traceless $\mathbf{14}$%
, and the rank-$3$ antisymmetric skew-traceless $\mathbf{14}^{\prime }$;
this latter characterizes $Sp(6,\mathbb{R})$ as a group \textquotedblleft of
type $E_{7}$" \cite{brown}.} $R=V(\lambda _{3})=:\mathbf{14}^{\prime }$,
namely the rank-$3$ completely antisymmetric skew-traceless representation,
which is an irreducible component of $\wedge ^{3}\mathbf{6=:6}_{a}^{\otimes
3}$ (where $\mathbf{6}$ is the fundamental representation). The dimension $%
\dim S_{(a^{p})}(\mathbf{14}^{\prime })^{Sp(6,\mathbb{R})}$ for the
partition $\lambda =a^{p}$, yielding the (real) dimension of the vector
space of polynomial invariants of homogeneity degree $pa$ in $p\dim R=14p$
variables, is given as above:%
\begin{equation}
\dim \left[ S_{\lambda =a^{p}}\left( \mathbf{14}^{\prime }\right) \right]
^{Sp(6,\mathbb{R})}=\dim \left[ S^{pa}\left( \mathbf{p},\mathbf{14}^{\prime
}\right) \right] ^{SL_{h}\left( p,\mathbb{R}\right) \times Sp(6,\mathbb{R}%
)}=:d.
\end{equation}

By perusing the first few $a$'s for the first few $p$'s, one gets the
following table:
\begin{equation}
\begin{array}{|c|c|c|c|c|c|c|c|c|c|c|c|c|}
\hline
Sp(6,\mathbb{R}),~\mathbf{14}^{\prime } & a= & 0 & 1 & 2 & 3 & 4 & 5 & 6 & 7
& 8 & 9 & 10 \\ \hline
p=2 & d= & 1 & 1 & 1 & 2 & 3 & 3 & 5 & 6 & 7 & 9 & 11 \\ \hline
p=3 & d= & 1 & 0 & 0 & 0 & 4 & 0 & 0 & 0 & 33 &  &  \\ \hline
p=4 & d= & 1 & 1 & 2 & 5 & 13 & 28 &  &  &  &  &  \\ \hline
p=5 & d= & 1 & 0 & 0 & 0 & 17 &  &  &  &  &  &  \\ \hline
p=6 & d= & 1 & 1 & 2 & 8 &  &  &  &  &  &  &  \\ \hline
p=7 & d= & 1 & 0 & 0 & 0 &  &  &  &  &  &  &  \\ \hline
p=8 & d= & 1 & 1 & 2 &  &  &  &  &  &  &  &  \\ \hline
\end{array}
\label{TableSp6}
\end{equation}%
Considerations essentially analogous to the ones made for the case of $%
G_{4}=E_{7}$ and $R=\mathbf{56}$ hold in this case, and in subsequent cases,
as well.

Note that the $p=2$ row of Table (\ref{TableSp6}) is identical to the $p=2$
row of Table (\ref{TableE7}); thus, the structure of the ring of polynomial
invariants of $\left( \mathbf{2},\mathbf{14}^{\prime }\right) $ of $SL_{h}(2,%
\mathbb{R})\times Sp(6,\mathbb{R})$ is the very same as the one of $\left(
\mathbf{2},\mathbf{56}\right) $ of $SL_{h}(2,\mathbb{R})\times E_{7}$. The
same will hold for all other examples of groups \textquotedblleft of type $%
E_{7}$" relevant to $D=4$ supergravity which we will consider below, meaning
that the structure of two-centered invariants, as well as their
interpretation (\ref{TableE7-p=2}), is the very same in all these cases.

However, this does not hold any more already starting from the $3$-centered
case ($p=3$), as it is immediate to realize by comparing the $p=3$ rows of (%
\ref{TableE7}) and (\ref{TableSp6}). Indeed, Table (\ref{TableSp6}), as
Table (\ref{TableE7}), yields that there are no $Sp(6,\mathbb{R})$%
-invariants for the partitions $\lambda =1^{3}$, $2^{3}$, $3^{3}$ and hence
there are no polynomial invariants of $\left( \mathbf{3},\mathbf{14}^{\prime
}\right) $ of $SL_{h}\left( 3,\mathbb{R}\right) \times Sp(6,\mathbb{R})$
with homogeneity degree $\leq 10$, the lowest possible degree being $12$, at
which however \texttt{Lie} finds $4$ invariants, instead of $5$ invariants
as in the $E_{7}$ case treated above. As above, the absence of an invariant
corresponding to the partition $\lambda =1^{3}$, \textit{i.e.} of a
\textquotedblleft $3$-centered analogue" of $\mathcal{W}$ (\ref{W}), can be
explained by the fact that $\mathbf{1}\notin \mathbf{14}_{a}^{\prime \otimes
3}$.


\subsubsection{{$G_{4}=SO(12)$}, $R=\mathbf{32}^{(\prime )}$}

This is related to the $D=4$ theories with symmetric scalar manifold, based
on the FTS's $\mathfrak{M}\left( J_{3}^{\mathbb{H}}\right) $ (magic $%
\mathcal{N}=2$ Maxwell-Einstein supergravity, sharing the same bosonic
sector of $\mathcal{N}=6$ supergravity, both with $G_{4}=SO^{\ast }(12)$
\cite{GST}) and $\mathfrak{M}\left( J_{3}^{\mathbb{H}_{s}}\right) $
(non-supersymmetric theory, with $G_{4}=SO(6,6)$ \cite{GBM}), where $J_{3}^{%
\mathbb{H}}$ and $J_{3}^{\mathbb{H}_{s}}$ are rank-$3$ Euclidean Jordan
algebras over the quaternions $\mathbb{H}$ and split quaternions $\mathbb{H}%
_{s}$, respectively.

In this case, the relevant {$SO(12)$}-representation is $R=\mathbf{32}$ or $%
R=\mathbf{32}^{\prime }$, namely one of the two chiral spinor
representations. The dimension $\dim S_{(a^{p})}(\mathbf{32}^{(\prime
)})^{SO(12)}$ for the partition $\lambda =a^{p}$, yielding the (real)
dimension of the vector space of polynomial invariants of homogeneity degree
$pa$ in $p\dim R=32p$ variables, is given as above:%
\begin{equation}
\dim \left[ S_{\lambda =a^{p}}\left( \mathbf{32}^{(\prime )}\right) \right]
^{SO(12)}=\dim \left[ S^{pa}\left( \mathbf{p},\mathbf{32}^{(\prime )}\right) %
\right] ^{SL_{h}\left( p,\mathbb{R}\right) \times SO(12)}=:d.
\end{equation}

By perusing the first few $a$'s for the first few $p$'s, one gets the
following table:
\begin{equation}
\begin{array}{|c|c|c|c|c|c|c|c|c|c|c|c|c|}
\hline
SO(12),~\mathbf{32}^{(\prime )} & a= & 0 & 1 & 2 & 3 & 4 & 5 & 6 & 7 & 8 & 9
& 10 \\ \hline
p=2 & d= & 1 & 1 & 1 & 2 & 3 & 3 & 5 & 6 & 7 & 9 & 11 \\ \hline
p=3 & d= & 1 & 0 & 0 & 0 & 5 & 0 & 0 & 0 &  &  &  \\ \hline
p=4 & d= & 1 & 1 & 2 & 5 & 17 & 42 &  &  &  &  &  \\ \hline
p=5 & d= & 1 & 0 & 0 & 0 & 42 &  &  &  &  &  &  \\ \hline
p=6 & d= & 1 & 1 & 3 & 14 &  &  &  &  &  &  &  \\ \hline
p=7 & d= & 1 & 0 & 0 &  &  &  &  &  &  &  &  \\ \hline
p=8 & d= & 1 & 1 & 4 &  &  &  &  &  &  &  &  \\ \hline
\end{array}
\label{TableSO(12)}
\end{equation}%
Considerations essentially analogous to the ones made for the cases of $%
G_{4}=E_{7}$, $R=\mathbf{56}$ and $G_{4}=Sp(6,\mathbb{R})$, $R=\mathbf{14}%
^{\prime }$ hold in this case, as well.


\subsubsection{{$G_{4}=SU(6)$}, $R=\mathbf{20}$}

This is related to the $D=4$ theories with symmetric scalar manifold, based
on the FTS's $\mathfrak{M}\left( J_{3}^{\mathbb{C}}\right) $ (magic $%
\mathcal{N}=2$ Maxwell-Einstein theory over $J_{3}^{\mathbb{C}}$, with $%
G_{4}=SU(3,3)$ \cite{GST}) and $\mathfrak{M}\left( J_{3}^{\mathbb{C}%
_{s}}\right) $ (non-supersymmetric theory, with $G_{4}=SL(6,\mathbb{R})$
\cite{GBM}), where $J_{3}^{\mathbb{C}}$ and $J_{3}^{\mathbb{C}_{s}}$ are
rank-$3$ Euclidean Jordan algebras over the complex numbers $\mathbb{C}$ and
split complex numbers $\mathbb{C}_{s}$, respectively\footnote{%
Actually, another supergravity theory exists in which $R=\mathbf{20}$,
namely $\mathcal{N}=5$, $D=4$ supergravity, with $U$-duality group $%
G_{4}=SU(1,5)$. However, this theory cannot be uplifted to $D=5$, and it is
not related to a FTS, but rather to the \textit{Jordan triple system} of $%
1\times 2$ octonionic vectors $M_{1,2}\left( \mathbb{O}\right) $ (see
\textit{e.g.} \cite{GST}, and Refs. therein).}.

In this case, the relevant $SU(6)$-representation is $R=\wedge ^{3}\mathbf{6}%
=:\mathbf{20}$, namely the rank-$3$ completely antisymmetric representation,
built out from the fundamental representation $\mathbf{6}$. Due to the
existence of the invariant $\epsilon $-tensor in the $\mathbf{6}$ of $SU(6)$%
, the irrep. $\mathbf{20}$ is real. The dimension $\dim S_{(a^{p})}(\mathbf{%
20})^{SU(6)}$ for the partition $\lambda =a^{p}$, yielding the (real)
dimension of the vector space of polynomial invariants of homogeneity degree
$pa$ in $p\dim R=20p$ variables, is given as above:%
\begin{equation}
\dim \left[ S_{\lambda =a^{p}}\left( \mathbf{20}\right) \right]
^{SU(6)}=\dim \left[ S^{pa}\left( \mathbf{p},\mathbf{20}\right) \right]
^{SL_{h}\left( p,\mathbb{R}\right) \times SU(6)}=:d.
\end{equation}

By perusing the first few $a$'s for the first few $p$'s, one gets the
following table:
\begin{equation}
\begin{array}{|c|c|c|c|c|c|c|c|c|c|c|c|c|}
\hline
SU(6),~\mathbf{20} & a= & 0 & 1 & 2 & 3 & 4 & 5 & 6 & 7 & 8 & 9 & 10 \\
\hline
p=2 & d= & 1 & 1 & 1 & 2 & 3 & 3 & 5 & 6 & 7 & 9 & 11 \\ \hline
p=3 & d= & 1 & 0 & 1 & 0 & 5 & 0 & 9 &  &  &  &  \\ \hline
p=4 & d= & 1 & 1 & 2 & 5 & 16 & 41 &  &  &  &  &  \\ \hline
p=5 & d= & 1 & 0 & 1 & 0 & 37 &  &  &  &  &  &  \\ \hline
p=6 & d= & 1 & 1 & 3 & 13 &  &  &  &  &  &  &  \\ \hline
p=7 & d= & 1 & 0 & 2 &  &  &  &  &  &  &  &  \\ \hline
p=8 & d= & 1 & 1 & 3 &  &  &  &  &  &  &  &  \\ \hline
\end{array}
\label{TableSU(6)}
\end{equation}%
Considerations essentially analogous to the previous cases hold in this
case, as well.


\subsubsection{\label{STU-gen}{$G_{4}=SL(2,\mathbb{R})\times SL(2,\mathbb{R}%
)\times SL(2,\mathbb{R})$}, $R=\left( \mathbf{2,2,2}\right) $}

We now consider the so-called $\mathcal{N}=2$ $stu$ model \cite{STU}, whose $%
U$-duality group is {$G_{4}=SL(2,\mathbb{R})\times SO(2,2)\cong SL(2,\mathbb{%
R})^{3}$}, with the relevant BH flux representation being the
tri-fundamental $R=\left( \mathbf{2},\mathbf{2},\mathbf{2}\right) $.

This provides an example of group \textquotedblleft of type $E_{7}$" \cite%
{brown} different from the ones treated above. Indeed, $SL(2,\mathbb{R})^{3}$
can still be characterized as a conformal symmetry, but of a \textit{%
semi-simple}, rank-$3$ Jordan algebra, namely $J_{3}=\mathbb{R}\oplus
\mathbb{R}\oplus \mathbb{R}$, or equivalently as the automorphism group of
the FTS $\mathfrak{M}\left( J_{3}\right) $ constructed over such an algebra:%
\begin{equation}
SL(2,\mathbb{R})^{3}=Conf\left( \mathbb{R}\oplus \mathbb{R}\oplus \mathbb{R}%
\right) =Aut\left( \mathfrak{M}\left( \mathbb{R}\oplus \mathbb{R}\oplus
\mathbb{R}\right) \right) .
\end{equation}%
Actually, by virtue of the isomorphism $\mathbb{R}\oplus \mathbb{R}\oplus
\mathbb{R\sim R}\oplus \Gamma _{1,1}$, this case can be regarded as the $%
\left( m,n\right) =\left( 2,2\right) $ element of the infinite sequence of
\textit{semi-simple} rank-$3$ Jordan algebras $\mathbb{R}\oplus \Gamma
_{m-1,n-1}$, where $\Gamma _{m-1,n-1}$ denotes the Clifford algebra of $%
O\left( m-1,n-1\right) $ \cite{JWVN}. This sequence can be related to $D=4$
supergravity theories (displaying symmetric scalar manifolds) for $m$(or
equivalently $n$)$=2$ ($\mathcal{N}=2$) or $6$ ($\mathcal{N}=4$). A complete
basis of minimal degree (which turns out to be \textit{finitely generating}
\cite{Kac}) of $2$-centered BH invariant polynomials have been firstly
determined in \cite{FMOSY-1}, and then further analyzed in \cite%
{CFMY-Small-1} and \cite{FMY-CV-1}.

The dimension $\dim S_{(a^{p})}(\left( \mathbf{2},\mathbf{2},\mathbf{2}%
\right) )^{SL(2,\mathbb{R})^{3}}$ for the partition $\lambda =a^{p}$,
yielding the (real) dimension of the vector space of polynomial invariants
of homogeneity degree $pa$ in $p\dim R=8p$ variables, is given as above:%
\begin{equation}
\dim \left[ S_{\lambda =a^{p}}\left( \left( \mathbf{2},\mathbf{2},\mathbf{2}%
\right) \right) \right] ^{SL(2,\mathbb{R})^{3}}=\dim \left[ S^{pa}\left(
\mathbf{p},\mathbf{2},\mathbf{2},\mathbf{2}\right) \right] ^{SL_{h}\left( p,%
\mathbb{R}\right) \times SL(2,\mathbb{R})^{3}}=:d.
\end{equation}

As done above, by perusing the first few $a$'s for the first few $p$'s, one
gets the following table:
\begin{equation}
\begin{array}{|c|c|c|c|c|c|c|c|c|c|c|c|c|}
\hline
SL(2,\mathbb{R})^{3},~\left( \mathbf{2},\mathbf{2},\mathbf{2}\right) & a= & 0
& 1 & 2 & 3 & 4 & 5 & 6 & 7 & 8 & 9 & 10 \\ \hline
p=2 & d= & 1 & 1 & 3 & 4 & 7 & 9 & 14 & 17 & 24 & 29 & 38 \\ \hline
p=3 & d= & 1 & 0 & 0 & 0 & 10 & 0 & 1 & 0 & 57 & 0 & 28 \\ \hline
p=4 & d= & 1 & 1 & 4 & 8 & 15 & 27 &  &  &  &  &  \\ \hline
p=5 & d= & 1 & 0 & 0 & 0 & 10 &  &  &  &  &  &  \\ \hline
p=6 & d= & 1 & 1 & 3 & 4 &  &  &  &  &  &  &  \\ \hline
p=7 & d= & 1 & 0 & 0 &  &  &  &  &  &  &  &  \\ \hline
p=8 & d= & 1 & 1 & 1 &  &  &  &  &  &  &  &  \\ \hline
\end{array}
\label{tab:stu}
\end{equation}%
We observe that the $p=2$ row of Table (\ref{tab:stu}) differs from the one
of Tables (\ref{TableE7}), (\ref{TableSp6}), (\ref{TableSO(12)}), (\ref%
{TableSU(6)}), which instead all share the same row. This can be traced back
to the \textit{semi-simple} nature of the rank-$3$ Jordan algebra $\mathbb{R}%
\oplus \Gamma _{1,1}$ to which the $stu$ model is be related, to be
contrasted to the \textit{simple} rank-$3$ Jordan algebras corresponding to
the cases treated above.

Moreover, it should be stressed that Table (\ref{tab:stu}) does not
implement a peculiar symmetry of the $stu$ model, namely the \textit{%
triality symmetry}\footnote{%
The relevance of this symmetry to the theory of \textit{Quantum Information}%
, and in particular to the classification of the quantum entanglement of
three (and four) qubits has been recently studied, exploiting techniques and
results from the supergravity side, also in the context of the so-called
\textit{BH/qubit correspondence} \cite{Duff-Cayley,QI-Refs,KL-QI}.},
corresponding to the invariance under the exchange of the three fundamentals
$\mathbf{2}$'s in $R=\left( \mathbf{2},\mathbf{2},\mathbf{2}\right) $,
achieved by imposing an invariance under the symmetric group $S_{3}$ acting
on the three $\mathbf{2}$'s in $R$.

The implementation of the \textit{triality symmetry} will be explicitly
worked out in Sec.\ \ref{stu} for the case of $p=3$ and $a=4$, namely for
the vector space of $3$-centered invariant polynomials of degree $12$,
which, from Table (\ref{tab:stu}), has dimension $10$; as yielded by the
treatment of Sec.\ \ref{co10in}, the dimension of the vector space of $3$%
-centered invariant polynomials of degree $12$ which are \textit{triality}-
(namely, $S_{3}$-) symmetric, and thus relevant for black holes in the $stu$
model, is $4$.

Our analysis can be refined as follows : by looking directly for the $\left(
SL_{h}(2,\mathbb{R})\times G_{4}\right) $-invariants as above, we now
consider the $G_{4}$-invariants in $S^{k}((\mathbb{R}^{2})\otimes R)$. The
formula (\ref{decomposition}) shows that these coincide with the $G_{4}$%
-invariants in $S_{\lambda }(R)$, tensored by the $SL_{h}(2,\mathbb{R})$%
-representation $S_{\lambda }(\mathbb{R}^{2})$, where $\lambda \vdash k$ and
$ht(\lambda )\leq 2$. By specifying this for the $stu$ model, as done in all
cases above, in \texttt{Lie} one types, for the partition $k=a+b$ with $%
a\geq b$, the following command (\textit{cfr.} \textit{e.g.} (\ref%
{nBHE7-input}))
\begin{equation}
\mbox{plethysm([$a$,$b$],[1,1,1],A1A1A1)[1]}.  \label{nBH-stu}
\end{equation}%
As mentioned, if $dX[0,0,0]$ occurs in the output, the coefficient $d$
yields the dimension of the space of $G_{4}$-invariants in $S_{(a,b)}(R)$,
otherwise there are no invariants in this representation.

In the $2$-centered case ($p=2$), an $S_{3}$-symmetric analysis of $SL(2,%
\mathbb{R})^{3}$- and $\left( SL_{h}(2,\mathbb{R})\times SL(2,\mathbb{R}%
)^{3}\right) $- invariant homogeneous polynomials for $2$-centered BHs in
the $stu$ model has been performed in \cite{FMOSY-1,CFMY-Small-1,FMY-CV-1}%
,whereas an $S_{4}$-symmetric treatment consistent in connection with the
quantum entanglement of four qubits was given in \cite{Levay-2-ctr}.

Indeed, the relevant $2$-centered representation for $stu$ model is actually
a \textit{quadri-fundamental} : for $p=2$ centers, one considers the
invariants of the group $SL_{h}(2,\mathbb{R})\times SL(2,\mathbb{R})^{3}$ in
the representation $\left( \mathbf{2},\mathbf{2},\mathbf{2},\mathbf{2}%
\right) $. Thus, one may promote the $S_{3}$-invariance (\textit{triality})
to an invariance (\textit{tetrality}) under the symmetric group $S_{4}$
acting on the four fundamentals $\mathbf{2}$'s in $\left( \mathbf{2},\mathbf{%
2},\mathbf{2},\mathbf{2}\right) $. A complete, minimal degree basis for the
ring of $\left( SL_{h}(2,\mathbb{R})\times SL(2,\mathbb{R})^{3}\right) $-
invariant homogeneous polynomials is given by $\mathcal{W}$, together with $%
2 $ quartic polynomials and with a sextic one, denoted by\footnote{%
Indeed, there is a slight difference in the definition of the $\left(
SL_{h}\left( 2,\mathbb{R}\right) \times G_{4}\right) $-invariant $\mathbf{I}%
_{6}$ for the models of $D=4$ (super)gravity based on \textit{simple} $J_{3}$%
's \cite{ADFMT-1} with respect to the definition of $\left( SL_{h}\left( 2,%
\mathbb{R}\right) \times G_{4}\right) $-invariant $\mathbf{I}_{6}^{\prime }$
for the models of $D=4$ (super)gravity based on the \textit{semi-simple}
sequence $J_{3,m,n}:=\mathbb{R}\oplus \mathbf{\Gamma }_{m-1,n-1}$ \cite%
{FMOSY-1,FMY-CV-1}; this is discussed in Sec.\ 3 of \cite{CFMY-Small-1}.} $%
\mathbf{I}_{6}^{\prime }$ \cite{Levay-2-ctr}.

When considering $2$-centered BH physics, one must discriminate between the
\textquotedblleft horizontal" symmetry $SL_{h}(2,\mathbb{R})$ \cite{FMOSY-1}
and the $U$-duality symmetry $G_{4}=SL(2,\mathbb{R})^{3}$, on which a
\textit{triality} must be implemented. Therefore, by down-grading $S_{4}$
(pertaining to four qubits in QIT) to $S_{3}$ (pertaining to $2$-centered $%
stu$ BHs), the consistent $S_{3}$-invariant $p=2$ counting performed in \cite%
{FMOSY-1,CFMY-Small-1,FMY-CV-1} yields that an invariant polynomial of
degree $8$ is no more generated by the previous ones, and a \textit{finitely
generating} \cite{Kac} complete basis for the ring of $\left( SL_{h}(2,%
\mathbb{R})\times SL(2,\mathbb{R})^{3}\right) $- invariant homogeneous
polynomials is given by four elements of degree $2$, $4$, $6$ and $8$ \cite%
{FMOSY-1}.

\section{\label{geometric}Geometric Interpretation}

In this section we consider the invariants for $SL_{h}(p)\times G_{4}$ in $(%
\mathbb{R}^{p})\otimes R=:\left( \mathbf{p},R\right) $ in the case that%
\footnote{%
In the case $p>r$, one can easily show that there are no non-trivial
invariants. This can be realized \textit{e.g.} as follows.
\par
One can write a tensor $t$ as $t=\sum_{i=1}^{p}f_{i}\otimes r_{i}$ (see Eq.\
(\ref{t})). In the case $p>r$, it is however more convenient to choose a
basis $e_{1},...,e_{r}$ of $R$, so that the same tensor can be rewritten as $%
t=\sum_{j=1}^{r}v_{j}\otimes e_{j}$, for (uniquely determined) vectors $%
v_{j}\in \mathbb{R}^{p}$.
\par
For a generic $t$ (to be precise, for $t$ outside the closed subset of
codimension $>1$ of $\mathbb{R}^{p}\otimes R$ defined by the vanishing of $%
r\times r$ minors of the matrix with rows $v_{1},...,v_{r}$), the vectors $%
v_{1},...,v_{r}$ are \textit{linearly independent}. Thus, there exists an
element $A\in SL_{h}(p,\mathbb{R})$ such that $Av_{i}=f_{i}$, where $\left\{
f_{i}\right\} $ is the standard basis of $\mathbb{R}^{p}$. Therefore, under
the action of $SL_{h}(p,\mathbb{R})\times \left\{ I\right\} $ all $t$'s in a
dense open subset of $\mathbb{R}^{p}\otimes R$ can be transformed into the
`standard' tensor $t=\sum_{j=1}^{r}f_{j}\otimes e_{j}$.
\par
Consequently, there is only one orbit (on this dense open set); as any $%
\left( SL_{h}(p,\mathbb{R})\times G_{4}\right) $-invariant polynomial must
be constant on this orbit, such a polynomial must be a constant, and thus
trivial. Note that in the limit case $r=p$, it could actually be given by
the determinant of the matrix $(v_{1},...,v_{p})$ (this is actually the
unique invariant in the case $r=p$), but if $r<p$ then the codimension of
the complement of this open orbit is $>1$, so a non-constant polynomial
would be zero in one point and non-zero in another point of the open orbit,
which yields a contradiction.}
\begin{equation}
p\,\leq \,r\,:=\,\dim R.  \label{ass-1}
\end{equation}%
Note that $r$ is even whenever the symplectic invariant $2$-form $\mathbb{C}%
_{MN}$ in $R_{a}^{\otimes 2}$ is non-degenerate (as we assume throughout the
paper).

We start and recall some classical results (mainly referring to \cite%
{Procesi}), and then we discuss the associated geometrical interpretation in
terms of Grassmannians.

The main result is the observation that the $G_{4}$-representation $%
S_{(a^{p})}(R)$ which, as discussed in Sec.\ \ref{algebraic}, produces all
invariants in $S^{ap}((\mathbb{R}^{p})\otimes R)$, can be identified with
the representation of $G_{4}$ on the homogeneous polynomials of degree $a$
in the \textit{Pl\"{u}cker coordinates} of the $p$-planes in $R$. Each of
these Pl\"{u}cker coordinates is an $SL_{h}(p)$-invariant homogeneous
polynomial of degree $p$ in the $p\dim R=pr$ coordinates on $(\mathbb{R}%
^{p})\otimes R$. Thus, the $G_{4}$-invariant polynomials homogeneous of
degree $a$ in these Pl\"{u}cker coordinates provide exactly the $\left(
SL_{h}(p)\times G_{4}\right) $-invariant homogeneous polynomials of degree $%
ap$ which are the object of our investigation.


\subsection{\label{grass}Grassmannians}


\subsubsection{\label{invsl}Invariants of $SL_{h}(p)\times G_{4}$ in $(%
\mathbb{R}^{p})\otimes R$}

Any tensor $t$ in $(\mathbb{R}^{p})\otimes R$ can be written as a sum $%
t=\sum_{a=1}^{\min \left( r,p\right) }x_{a}\otimes y_{a}$, with $x_{a}\in
\mathbb{R}^{p}$, $y_{a}\in R$. Let $f_{1},\ldots ,f_{p}$ be the standard
basis of $\mathbb{R}^{p}$. Writing each $x_{a}=\sum_{i=1}^{p}x_{ai}f_{i}$,
and using the bilinearity of $\otimes $, one finds that%
\begin{equation}
t=\sum_{i=1}^{p}f_{i}\otimes r_{i},  \label{t}
\end{equation}%
for certain uniquely determined elements $r_{i}\in R$.

Since any $\left( SL_{h}(p)\times G_{4}\right) $-invariant $F$ is obviously
an $\left( SL_{h}(p)\times \{I\}\right) $-invariant, it is firstly
convenient to study the invariants of $SL_{h}(p)\times \{I\}$. To this end,
we only consider the action of $SL_{h}(p)$ on the first factor of $(\mathbb{R%
}^{p})\otimes R$, so we are actually dealing with the direct sum of $r$
copies of the fundamental representation $\mathbb{R}^{p}=:\mathbf{p}$ of $%
SL_{h}(p)$. In the case $r\geq p$ (\ref{ass-1}), the ring of invariants in
this case is well understood. Fixing a basis $e_{1},\ldots ,e_{r}$ of $R$,
this ring is generated by the determinants of the $\left( p\times p\right) $%
-minors of the $r\times p$ matrix $T:=T_{t}$ whose columns are the vectors $%
r_{1},\ldots ,r_{p}$ (\cite{Procesi}, 11.1.2).

Note that all invariants $F$ vanish on the tensors $t=\sum_{i=1}^{p}f_{i}%
\otimes r_{i}$ such that the rank of the matrix $T_{t}$ is less than $p$,
\textit{i.e.}\ when the $r_{i}$ do not span a $p$-dimensional subspace of $R$%
; such tensors $t$ are called \textit{unstable} (\textit{i.e.},\ not
semi-stable) tensors for this action. The (geometric) quotient $((\mathbb{R}%
^{p})\otimes R)/\!/SL_{h}(p)$ is the image of the \textit{quotient map} $\pi
$ given by generators of the ring of invariants $F$ (\cite{Procesi},
11.1.2):
\begin{equation}
\pi :\,(\mathbb{R}^{p})\otimes R\,\longrightarrow \,\wedge ^{p}R,\qquad
t\,=\,\sum_{i=1}^{p}f_{i}\otimes r_{i}\,\longmapsto \,r_{1}\wedge
r_{2}\wedge \ldots \wedge r_{p}.  \label{pi}
\end{equation}%
Note that $\wedge ^{p}R=:R_{a}^{\otimes p}=S_{\lambda }\left( R\right) $
(with partition $\lambda =1^{p}$) has basis $e_{I}=e_{i_{1}}\wedge \ldots
\wedge e_{i_{p}}$, with $i_{1}<\ldots <i_{p}$, and therefore $\pi (t)=\sum
t_{I}e_{I}$ (with $I$ collectively denoting the indices $i_{1}<\ldots <i_{p}$%
), where $t_{I}$ is the determinant of the minor of $T_{t}$ formed by the
rows $i_{1},\ldots ,i_{p}$.

The image of the quotient map $\pi $ (\ref{pi}) consists of the decomposable
tensors in $\wedge ^{p}R$. This map, when restricted to stable points, is
the lift to linear spaces of the \textit{Pl\"{u}cker map} $%
Gr(p,R)\rightarrow \mathbf{P}(\wedge ^{p}R)$, where $Gr(p,R)$ denotes the
\textit{Grassmannian} of $p$-planes in $R$ (see Sec.\ \ref{pluck}).

Let now $F$ be an $\left( SL_{h}(p)\times G_{4}\right) $-invariant. Since it
is trivially an $\left( SL_{h}(p)\times \{I\}\right) $-invariant, from the
above reasoning $F$ is a polynomial in the determinants of $\left( p\times
p\right) $-minors of $T_{t}$. Therefore, all such invariants can be
determined with a two-step approach\footnote{%
It is funny to note that this approach is actually the opposite of the
method which has been exploited in supergravity (especially in the $2$%
-centered case $p=2$) : in that framework, the $G_{4}$-invariants are
organized in irreps. of $SL_{h}(p)$, from which one picks out the trivial
(singlet) $SL_{h}(p)$-representations (see \textit{e.g.} \cite%
{FMOSY-1,ADFMT-1,CFMY-Small-1,FMY-CV-1}).} :

\textbf{1}] first, one identifies the space of such polynomials as a
representation of $G_{4}$;

\textbf{2}] then, one finds the $G_{4}$-invariants in that space.

Step \textbf{1} is actually well-known when one considers the space of such
polynomials as a representation for the larger group $GL(R)=:GL(r)$ (namely,
within (\ref{ass-1})) : as a $GL(R)$-representation, the space of
polynomials, homogeneous of degree $a$ in the $\left( p\times p\right) $%
-minors of the $p\times r$ matrices, is $S_{a^{p}}(R)$ (\cite{Procesi},
11.1.2).

In order to find the $\left( SL_{h}(p)\times G_{4}\right) $-invariants in $(%
\mathbb{R}^{p})\otimes R$, it then suffices to find the $G_{4}$-invariants
in the representations $S_{a^{p}}(R)$ (step \textbf{2}). This conclusion was
already reached in Sec.\ \ref{p-bH}; however, the above discussion clarifies
how a $G_{4}$-invariant in $S_{a^{p}}(R)$ produces a polynomial on $(\mathbb{%
R}^{p})\otimes R$.

We are now going to reformulate this reasoning in a geometrical way.


\subsubsection{From Tensors to Planes}

In order to study $p$-centered BHs, for the case (\ref{ass-1}), one can use
the Grassmannian $Gr(p,R)$ of $p$-planes in $R$ as follows.

Using the notation of Sec.\ \ref{invsl}, any tensor $t$ in $(\mathbb{R}%
^{p})\otimes R$ can be written as $t=\sum_{i=1}^{p}f_{i}\otimes r_{i}$, for
certain uniquely determined elements $r_{i}\in R$. It is here convenient to
consider the dense open subset
\begin{equation}
(\mathbb{R}^{p}\otimes R)^{0}\,:=\,\left\{ \sum_{i=1}^{p}f_{i}\otimes
r_{i}:\;\dim \langle r_{1},\ldots ,r_{p}\rangle \,=\,p\,\right\} ,
\end{equation}%
such that the $p$ vectors $r_{1},\ldots ,r_{p}$ span a $p$-dimensional
subspace of $R$ (the upperscript \textquotedblleft $0$" denotes the absence
of unstable points). This yields a map $\mathbf{G}$ to $Gr(p,R)$ as follows:
\begin{equation}
\mathbf{G}:\,((\mathbb{R}^{p})\otimes R)^{0}\,\longrightarrow
\,Gr(p,R),\qquad t\,=\,\sum_{i=1}^{p}f_{i}\otimes r_{i}\,\longmapsto
\,W_{t}\,:=\,\langle r_{1},\ldots ,r_{p}\rangle .  \label{G-map}
\end{equation}%
It is worth noting that the action of $SL_{h}(p)$ on $\mathbb{R}^{p}$ merely
changes the basis of $W_{t}$, so the map $\mathbf{G}$ is $SL_{h}(p)$%
-invariant. It is obviously also $GL_{h}(p)$-invariant, so it is actually
identifying more tensors than strictly necessary for our purposes. The map $%
\mathbf{G}$ (\ref{G-map}), besides being injective, is obviously also
surjective: indeed, given a $p$-plane $W\subset R$, one can choose a basis $%
r_{1},\ldots ,r_{p}$, and then $W=W_{t}$, where $t=\sum_{i=1}^{p}f_{i}%
\otimes r_{i}$. Thus, one gets the following bijection
\begin{equation}
((\mathbb{R}^{p})\otimes R)^{0}/GL_{h}(p)\;\longleftrightarrow
\;Gr(p,R),\qquad t\,\longleftrightarrow \,W_{t}.  \label{bijection}
\end{equation}%
In particular, any $G_{4}$-invariant function on the Grassmannian $Gr(p,R)$
of $p$-planes in $R$ will yield an $\left( SL_{h}(p)\times G\right) $%
-invariant function on $((\mathbb{R}^{p})\otimes R)^{0}$, which will
eventually extend\footnote{%
In the present investigation, as resulting from Sec.\ \ref{algebraic}, we
consider\textit{\ homogeneous polynomial} invariants; in such a case, the
extension from $((\mathbb{R}^{p})\otimes R)^{0}$ to the whole $(\mathbb{R}%
^{p})\otimes R$ is immediate.} to the whole relevant irrep. $(\mathbb{R}%
^{p})\otimes R$.


\subsubsection{\label{pluck}The Pl\"{u}cker Map}

As $Gr(p,R)$ is (a real subset of) a projective variety, which is moreover a
$p\left( r-p\right) $-dimensional homogeneous space:%
\begin{equation}
Gr\left( p,r\right) \cong \frac{O\left( r\right) }{O\left( p\right) \otimes
O\left( r-p\right) },
\end{equation}%
one can proceed as follows. Recall that the \textit{Pl\"{u}cker map} $%
\mathcal{P}$ is defined as the embedding
\begin{equation}
\mathcal{P}:Gr(p,R)\,\longrightarrow \,\mathbf{P}(\wedge ^{p}R),\qquad
W_{t}\,\longmapsto \,\wedge ^{p}W_{t}.  \label{P-map}
\end{equation}%
In particular, the composition $\mathcal{P}\circ \mathbf{G}$ of this map
with $\mathbf{G}$ (\ref{G-map}) maps $t$ to $r_{1}\wedge \ldots \wedge r_{p}$%
. Fixing a basis $e_{1},\ldots ,e_{r}$ of $R$, one thus gets the basis $%
e_{I}=e_{i_{1}}\wedge \ldots \wedge e_{i_{p}}$, with $i_{1}<\ldots <i_{p}$ ,
of $\wedge ^{p}R$ (\textit{cfr.} below (\ref{pi})). The \textit{Pl\"{u}cker
coordinates} of $W_{t}$ are defined as the $\left( p\times p\right) $-minors
of the $r\times p$ matrix $T:=T_{t}$ with columns $r_{1},\ldots ,r_{p}$.

The action of the group $GL_{h}(R)$ can be represented on the space of
global sections $\Gamma (Gr(p,R),L)$ on a line bundle $L$ over $Gr(p,R)$.
Working over the complex numbers and denoting by $Pic\left( X\right) $ the
\textit{Picard group} of the variety $X$, let us recall that $Pic\left(
Gr\left( p,R\right) \right) $ is generated by a (very ample) line bundle $L$%
, whose global sections are the Pl\"{u}cker coordinates themselves. In fact,
$\Gamma (Gr(p,R),L)\cong \wedge ^{p}R$, (actually the dual representation
thereof, since the coordinates are linear maps on $\wedge ^{p}R$). The
action of $GL_{h}(R)$ on $R$ then induces an action on the Grassmannian $%
Gr(p,R)$ and thus on the spaces of global sections $\Gamma (Gr(p,R),L)$. By
recalling that $\wedge ^{p}R=S_{\lambda }\left( R\right) $ with partition $%
\lambda =1^{p}$ (\textit{cfr.} below (\ref{dim-S-lambda})), Bott's theorem
(see \textit{e.g.} \cite{Bott-Th}) gives, as $GL_{h}(R)$-representations:
\begin{equation}
\Gamma (Gr(p,R),\,L^{\otimes a})\,\cong \,S_{a\lambda }(R),\qquad a\lambda
\,:=\underbrace{(a,\ldots ,a)}_{p}\,=:\,a^{p}.  \label{Bott-Th}
\end{equation}%
Furthermore, any global section of $L^{\otimes a}$ is a linear combination
of products of $a$ sections of $L$ (and therefore the map $S^{a}\Gamma
(L)\rightarrow \Gamma (L^{\otimes a})$ is surjective); in terms of
representations, this simply amounts to the statement that $S_{a\lambda }$
is a summand of $S^{ap}(R)$. Thus, any section of $L^{\otimes a}$ is a
homogeneous polynomial in the Pl\"{u}cker coordinates of degree $a$.

Given a $G_{4}$-invariant $F\in S_{a\lambda }(R)\cong \Gamma
(Gr(p,R),L^{\otimes a})$, it corresponds to a degree $a$ homogeneous
polynomial in the Pl\"{u}cker coordinates, defined by the map (recall (\ref%
{G-map}) and (\ref{bijection})):%
\begin{equation}
F:Gr(p,R)\,\longrightarrow \,\mathbb{R},  \label{F-map}
\end{equation}%
Thus, the composition
\begin{equation}
F\circ \mathbf{G}:\,((\mathbb{R}^{p})\otimes R)^{0}\,\longrightarrow
\,\wedge ^{p}R\longrightarrow \mathbb{R},  \label{FG}
\end{equation}%
yields a $\left( SL_{h}(p)\times G_{4}\right) $-invariant which extends to
the whole $(\mathbb{R}^{p})\otimes R$. This provides a geometrical
explanation of the treatment of Sec.\ \ref{algebraic}, and in particular of
the fact that the $S_{\lambda }\left( R\right) $ with $\lambda =a^{p}$
contribute to - \textit{and actually are the unique responsible for} - the $%
\left( SL_{h}(p)\times G_{4}\right) $-invariant homogeneous polynomials in $(%
\mathbb{R}^{p})\otimes R$.\smallskip

To summarize, in order to find $\left( SL_{h}(p)\otimes G_{4}\right) $%
-invariant homogeneous polynomials $F$ in the representation $(\mathbb{R}%
^{p})\otimes R$, one needs to find invariant polynomials $\hat{F}$ for the
induced action of $G_{4}$ on $\wedge ^{p}R$:
\begin{equation}
F(t)\,=\,\hat{F}(\ldots ,p_{i_{1}\ldots i_{p}}(t),\ldots ),  \label{FF}
\end{equation}%
where $p_{i_{1}\ldots i_{p}}(t)=p_{\left[ i_{1}\ldots i_{p}\right] }(t)$.

In particular, if an invariant $F$ is a homogeneous polynomial of degree $k$
in the coefficients $c_{ij}$ of $t=\sum c_{ij}f_{i}\otimes e_{j}$, then, as
each Pl\"{u}cker coordinate is homogeneous of degree $p$ in the $c_{ij}$, $%
\hat{F}$ is homogeneous of degree $k/p$ in the Pl\"{u}cker coordinates. Thus,%
\textit{\ }$k$\textit{\ must be a multiple of }$p$\textit{.} This matches
the statement made below (\ref{res-1}), and it is not surprising, as $SL(p,%
\mathbb{C})$ contains the diagonal matrices $\omega I$ where $\omega
=e^{2\pi i/p}$ and these act by multiplication by $\omega ^{d}$ on
polynomials $F$ of degree $k$; so, if $F$ is $SL_{h}(p)$-invariant, $k$ must
indeed be a multiple of $p$. Moreover, these invariants $\hat{F}$ should be
non-zero when restricted to the (semi-)stable decomposable tensors.


\section{\label{stu}$3$-centered $stu$ Black Holes}

We will now apply the method discussed in Secs. \ref{algebraic} and \ref%
{geometric} to compute the invariants pertaining to $3$-centered ($p=3$) BHs
in the $\mathcal{N}=2$, $D=4$ $stu$ model \cite{STU}. As discussed in Sec.\ %
\ref{STU-gen}, in this case the $U$-duality group is {$G_{4}=SL(2,\mathbb{R}%
)\times SO(2,2)\cong SL(2,\mathbb{R})^{3}$}, with the relevant BH
representation being the tri-fundamental $R=\left( \mathbf{2},\mathbf{2},%
\mathbf{2}\right) $. Moreover, the $K$-tensor (namely, the unique rank-$4$
symmetric invariant in $\left( \mathbf{2},\mathbf{2},\mathbf{2}\right)
_{s}^{\otimes 4}$; see Sec.\ \ref{Intro}) is given by the \textit{Cayley's
hyperdeterminant} on $R$ \cite{Cayley,Duff-Cayley}.

In Table (\ref{tab:stu}), we have computed the dimension of the spaces of
invariants for $p=3$ up to degree $30$. In particular, the lowest degree
non-trivial $\left( SL_{h}\left( 3,\mathbb{R}\right) \times SL\left( 2,%
\mathbb{R}\right) ^{3}\right) $-invariant homogeneous polynomials in the $%
\mathbb{R}^{3}\otimes \left( \mathbf{2},\mathbf{2},\mathbf{2}\right) =:(%
\mathbf{3},\mathbf{2},\mathbf{2},\mathbf{2})$ have degree $12$, and they
span a $10$-dimensional space.

From the treatment of Secs. \ref{algebraic} and \ref{geometric}, as well as
from Table (\ref{tab:stu}), such $3$-centered invariant polynomials lie in $%
S_{4^{3}}\left( \left( \mathbf{2},\mathbf{2},\mathbf{2}\right) \right) $. In
the present Section, we will determine a basis for their $10$-dimensional
space. Then, in Subsubsec.\ \ref{co10in} we will implement invariance (%
\textit{triality}) under the $S_{3}$ symmetric group acting on the three $%
\mathbf{2}$'s in $R$, obtaining a basis of the resulting $4$-dimensional
vector space of $\left( S_{3}\times SL_{h}\left( 3,\mathbb{R}\right) \times
SL\left( 2,\mathbb{R}\right) ^{3}\right) $-invariant homogeneous polynomials
of degree $12$ in the $(\mathbf{3},\mathbf{2},\mathbf{2},\mathbf{2})$, thus
pertaining to the description of $3$-centered BHs in the $stu$ model.


\subsection{\label{oci}Invariant from Cayley's Hyperdeterminant : $%
S^{4}(S_{1^{3}}\left( \left( \mathbf{2},\mathbf{2},\mathbf{2}\right) \right)
)$}

A first invariant can be constructed as follows.

Let us recall that the BH flux irrep. $R=\left( \mathbf{2},\mathbf{2},%
\mathbf{2}\right) $ is endowed with an invariant alternating form $\mathbb{C}
$, \textit{i.e.} the symplectic $8\times 8$ metric $\mathbb{C}%
_{MN}:=(\exists !)\mathbf{1}\in \left( \mathbf{2},\mathbf{2},\mathbf{2}%
\right) _{a}^{\otimes 2}$. Within the notation of Sec.\ \ref{grass}, the
restriction of $\mathbb{C}_{MN}$ to the $3$-dimensional subspace $%
W_{t}\subset \left( \mathbf{2},\mathbf{2},\mathbf{2}\right) $ generated by $%
3 $ given charge vectors $Q_{i}=:r_{i}\in \left( \mathbf{2},\mathbf{2},%
\mathbf{2}\right) $ (we here denote the \textquotedblleft horizontal" index
as $i=1,2,3=p$) is given by the $3\times 3$ alternating matrix
\begin{equation}
\mathbb{C}_{t}:=\left. \mathbb{C}\right\vert _{W_{t}\otimes W_{t}}\,=\,%
\begin{pmatrix}
0 & (r_{1},r_{2}) & (r_{1},r_{3}) \\
(r_{2},r_{1}) & 0 & (r_{2},r_{3}) \\
(r_{3},r_{1}) & (r_{3},r_{2}) & 0%
\end{pmatrix}%
,\qquad W_{t}\,:=\,\langle r_{1},r_{2},r_{3}\rangle \,\subset \,\left(
\mathbf{2},\mathbf{2},\mathbf{2}\right) ,  \label{Ct}
\end{equation}%
where (\textit{cfr.} (\ref{W}); $M=1,...,8=\dim \left( \mathbf{2},\mathbf{2},%
\mathbf{2}\right) $)%
\begin{equation}
\left( \mathbb{C}_{t}\right) _{ij}=(r_{i},r_{j}):=\mathbb{C}%
_{MN}r_{i}^{M}r_{j}^{N}=:\mathcal{W}_{ij}=-\mathcal{W}_{ji}
\end{equation}%
is the $SL\left( 2,\mathbb{R}\right) ^{3}$-invariant symplectic product of $%
r_{i}$ and $r_{j}$. It is immediate to realize that $\mathcal{W}_{ij}$ ($%
i,j=1,2,3$) belongs to the $\mathbf{3}^{\prime }=\wedge ^{2}\mathbf{3}$ of $%
SL_{h}\left( 3,\mathbb{R}\right) $ (\textit{cfr.} end of Sec.\ \ref{k=2}, as
well as the end of Sec.\ 3 of \cite{ADFMT-1}); indeed, by using the
Ricci-Levi-Civita invariant symbol $\epsilon ^{ijk}$ of $SL_{h}\left( 3,%
\mathbb{R}\right) $, one can define%
\begin{equation}
\mathcal{W}^{i}:=\frac{1}{2}\epsilon ^{ijk}\mathbb{C}_{MN}r_{j}^{M}r_{k}^{N}=%
\frac{1}{2}\epsilon ^{ijk}\mathcal{W}_{jk}\in \left( \mathbf{3}^{\prime },%
\mathbf{1,1,1}\right) ~\text{of~}SL_{h}\left( 3,\mathbb{R}\right) \times
SL\left( 2,\mathbb{R}\right) ^{3}.
\end{equation}

The vector
\begin{equation}
v_{t}\,:=\,(r_{2},r_{3})r_{1}\,+\,(r_{3},r_{1})r_{2}\,+\,(r_{1},r_{2})r_{3}=%
\frac{1}{2}\epsilon ^{ijk}\mathcal{W}_{jk}r_{i}=\frac{1}{2}\epsilon ^{ijk}%
\mathcal{W}_{[jk}r_{i]}\in W_{t}  \label{vt}
\end{equation}%
spans the kernel of $\mathbb{C}_{t}$ (\ref{Ct}), and it can be considered as
a multilinear alternating map
\begin{equation}
v_{t}\,:\left( \mathbf{2},\mathbf{2},\mathbf{2}\right) ^{\otimes
3}\,\longrightarrow \,\left( \mathbf{2},\mathbf{2},\mathbf{2}\right) ,\qquad
(r_{1},r_{2},r_{3})\,\longmapsto
\,(r_{2},r_{3})r_{1}\,+\,(r_{3},r_{1})r_{2}\,+\,(r_{1},r_{2})r_{3}.
\label{vt-map}
\end{equation}%
In order to see this, it suffices to check that it is alternating for the
permutations $(12)$ and $(23)$, which is easily done. Thus, the map $v_{t}$ (%
\ref{vt-map}) induces a linear map $\wedge ^{3}\left( \mathbf{2},\mathbf{2},%
\mathbf{2}\right) \rightarrow \left( \mathbf{2},\mathbf{2},\mathbf{2}\right)
$; by virtue of the treatment of Sec.\ \ref{geometric}, this proves that $%
v_{t}$ is a linear combination of the $r_{i}$ with coefficients which are
linear forms in the Pl\"{u}cker coordinates of $t$. From the treatment of
Sec.\ \ref{geometric}, these Pl\"{u}cker coordinates are homogeneous of
degree $p=3$ in the coordinates $c_{ij}$ of $t$, and they are invariant
under the action of $SL_{h}(3,\mathbb{R})$, hence%
\begin{equation}
v_{t}=v_{(A,I)t},~~\forall A\in SL_{h}(3,\mathbb{R}),  \label{rez-1}
\end{equation}%
implying that%
\begin{equation}
v_{t}\in \left( \mathbf{1},W_{t}\right) \subset \left( \mathbf{1},\mathbf{2},%
\mathbf{2},\mathbf{2}\right) .  \label{rez-2}
\end{equation}

As the symplectic $2$-form $\mathbb{C}$ is $SL\left( 2,\mathbb{R}\right)
^{3} $-invariant, by recalling definition (\ref{vt}) one obtains the
following formula for the action of $B\in SL\left( 2,\mathbb{R}\right) ^{3}$
on $v_{t}$ itself:{\
\begin{equation}
\begin{array}{rcl}
Bv_{t} & = & (r_{2},r_{3})Br_{1}\,+\,(r_{3},r_{1})Br_{2}\,+%
\,(r_{1},r_{2})Br_{3} \\
& = & (Br_{2},Br_{3})Br_{1}\,+(Br_{3},Br_{1})Br_{2}\,+\,(Br_{1},Br_{2})Br_{3}
\\
& = & v_{(I,B)t}.%
\end{array}
\label{rr}
\end{equation}%
} 
By virtue of (\ref{rez-1}), since%
\begin{equation}
v_{(A,B)t}=v_{(A,I)(I,B)t}=v_{(I,B)t}=Bv_{t},
\end{equation}%
any $SL\left( 2,\mathbb{R}\right) ^{3}$-invariant polynomial $F$ of degree $%
g $ on the tri-fundamental representation $R=(\mathbf{2},\mathbf{2},\mathbf{2%
}) $ produces an $\left( SL_{h}\left( 3,\mathbb{R}\right) \times SL\left( 2,%
\mathbb{R}\right) ^{3}\right) $-invariant polynomial $F_{0}$ homogeneous of
degree $3g$ on $\left( \mathbf{3},\mathbf{2},\mathbf{2},\mathbf{2}\right) $,
defined as follows:
\begin{equation}
F_{0}(t)\,:=\,F(v_{t}).
\end{equation}

A natural choice is $F=\mathcal{I}_{4}$, where $\mathcal{I}_{4}$ is the
\textit{Cayley's hyperdeterminant} \cite{Cayley} on $(\mathbf{2},\mathbf{2},%
\mathbf{2})$ (determined by the $K$-tensor of $(\mathbf{2},\mathbf{2},%
\mathbf{2})$ \cite{Cayley,Duff-Cayley}); this is an homogeneous polynomial
of degree $4$, and it is the \textit{unique} algebraically independent $%
SL\left( 2,\mathbb{R}\right) ^{3}$-invariant polynomial on the $(\mathbf{2},%
\mathbf{2},\mathbf{2})$ itself. Therefore, the choice $F=\mathcal{I}_{4}$
yields an $\left( SL_{h}\left( 3,\mathbb{R}\right) \times SL\left( 2,\mathbb{%
R}\right) ^{3}\right) $-invariant polynomial $F_{0}$ homogeneous of degree $%
3\cdot 4=12$ for $3$-centered BHs in the $stu$ model:%
\begin{equation}
F_{0}(t):=\mathcal{I}_{4}(v_{t}).  \label{Inv}
\end{equation}%
\medskip

The construction performed above can be clarified in terms of representation
theory as follows.

From the treatment of Secs. \ref{algebraic} and \ref{geometric} (in
particular, recalling (\ref{FF})), the $\left( SL_{h}(3,\mathbb{R})\otimes
G_{4}\right) $-invariants homogeneous polynomials $F$ on $(\mathbb{R}%
^{3})\otimes R=:\left( \mathbf{3},R\right) $ are given by invariants $\hat{F}
$ for the induced action of $G_{4}$ on $\wedge ^{3}R=:R_{a}^{\otimes
3}=S_{\lambda }(R)$ (with partition $\lambda :=1^{3}$; see below (\ref%
{dim-S-lambda})):
\begin{equation}
F(t)\,=\,\hat{F}(\ldots ,p_{i_{1}i_{2}i_{3}}(t),\ldots ),
\end{equation}%
(where $p_{i_{1}i_{2}i_{3}}(t)=p_{\left[ i_{1}i_{2}i_{3}\right] }(t)$) which
should be non-zero when restricted to the (semi-)stable\texttt{\ }%
decomposable tensors.

In general, the representations of $G_{4}$ on $\wedge ^{3}R$ may be
reducible. Indeed, for the $stu$ model we have
\begin{equation}
S_{1^{3}}(\left( \mathbf{2},\mathbf{2},\mathbf{2}\right) )\,:=\,\wedge
^{3}\left( \mathbf{2},\mathbf{2},\mathbf{2}\right) \,\equiv \left( \mathbf{2}%
,\mathbf{2},\mathbf{2}\right) _{a}^{\otimes 3}\cong \,\left( \mathbf{2},%
\mathbf{2},\mathbf{2}\right) \oplus \left( \mathbf{4},\mathbf{2},\mathbf{2}%
\right) \oplus \,\left( \mathbf{2},\mathbf{4},\mathbf{2}\right) \oplus
\,\left( \mathbf{2},\mathbf{2},\mathbf{4}\right) ,  \label{s111}
\end{equation}%
where $\mathbf{4}$ denotes the spin $s=3/2$ irrep. of $SL(2,\mathbb{R})$.

The appearance of $\left( \mathbf{2},\mathbf{2},\mathbf{2}\right) $ in the
r.h.s. of (\ref{s111}), and in general the fact that $R\in R_{a}^{\otimes 3}$%
, can be simply related to the existence of the $G_{4}$-equivariant map
\begin{equation}
R\,\longrightarrow \,\wedge ^{3}R,\qquad r\longmapsto \mathbb{C}^{\ast
}\wedge r
\end{equation}%
where $\mathbb{C}\in \wedge ^{2}R^{\ast }$ corresponds to $\mathbb{C}^{\ast
}\in \wedge ^{2}R$ under the duality given by the non-degenerate symplectic
form $\mathbb{C}\equiv \mathbb{C}_{MN}=$ $\mathbb{C}_{\left[ MN\right] }$ on
$R$ (symplectic structure of - generalized - electric-magnetic duality in $%
D=4$). This implies that any $G_{4}$-invariant on $R$ trivially produces a $%
G_{4}$-invariant on $\wedge ^{3}R$.

Let us call $\Psi _{t}$ the generalization (for a generic case) of the map $%
v_{t}$ (\ref{vt})-(\ref{vt-map}) constructed above:
\begin{equation}
\Psi _{t}:\,\wedge ^{3}R\,\longrightarrow \,R,\quad r_{1}\wedge r_{2}\wedge
r_{3}\,\longmapsto \,(r_{2},r_{3})r_{1}+(r_{3},r_{1})r_{2}+(r_{1},r_{2})r_{3}
\label{Psi}
\end{equation}%
which then satisfies (\textit{cfr.} (\ref{rr}))
\begin{equation}
\Psi _{t}(B(r_{1}\wedge r_{2}\wedge r_{3}))\,:=\Psi _{t}((Br_{1})\wedge
(Br_{2})\wedge (Br_{3}))\,=\,B\Psi _{t}(r_{1}\wedge r_{2}\wedge
r_{3}),~\forall B\in G_{4},
\end{equation}%
since $(Br_{i},Br_{j})=(r_{i},r_{j})$. Thus the map $\Psi _{t}$ (\ref{Psi})
is, up to scalar multiplication, the unique $G_{4}$-equivariant projection
of $\wedge ^{3}R$ onto $R$.

Thus, coming back to the $stu$ model, it follows that, up to a real scalar,
the map $\pi :\left( \mathbf{3},\mathbf{2},\mathbf{2},\mathbf{2}\right)
\rightarrow \wedge ^{3}\left( \mathbf{2},\mathbf{2},\mathbf{2}\right) $ (%
\textit{cfr.} (\ref{pi}) for $p=3$) is given by
\begin{equation}
\pi (t)\,=\,v_{t}+w_{t},\qquad v_{t}:=\Psi _{t}(W_{t})\in \left( \mathbf{2},%
\mathbf{2},\mathbf{2}\right) ,\quad w_{t}\in \left( \mathbf{4},\mathbf{2},%
\mathbf{2}\right) \oplus \,\left( \mathbf{2},\mathbf{4},\mathbf{2}\right)
\oplus \,\left( \mathbf{2},\mathbf{2},\mathbf{4}\right) .
\end{equation}%
This leads to the invariant $F_{0}$ (\ref{Inv}), which is thus given by the
image of $S^{4}(S_{1^{3}}\left( \mathbf{2},\mathbf{2},\mathbf{2}\right) )$
in $S_{4,4,4}\left( \left( \mathbf{2},\mathbf{2},\mathbf{2}\right) \right) $%
.\medskip

From the treatment above, it clearly follows that the degree-$12$
homogeneous $\left( SL_{h}(3,\mathbb{R})\times G_{4}\right) $-invariant
polynomial $F_{0}$ (\ref{Inv}) can be consistently defined for all groups $%
G_{4}$ \textquotedblleft of type $E_{7}$", and in particular \textit{at least%
} for the class relevant to $D=4$ supergravity theories with symmetric
scalar manifolds, listed in Table 1.

\subsection{\label{oci-2}Other Invariants from $S^{2}(S_{2^{3}}\left( \left(
\mathbf{2},\mathbf{2},\mathbf{2}\right) \right) )$}

As a natural next step, one can try to determine other $SL\left( 2,\mathbb{R}%
\right) ^{3}$-invariants of degree $12$ from quadratic invariants in $%
S_{2^{3}}\left( \left( \mathbf{2},\mathbf{2},\mathbf{2}\right) \right) $.

Using \texttt{LiE}, one can decompose $S_{2^{3}}\left( \left( \mathbf{2},%
\mathbf{2},\mathbf{2}\right) \right) $ into irreducible $SL\left( 2,\mathbb{R%
}\right) ^{3}$-representations:
\begin{equation}
S_{2^{3}}\left( \left( \mathbf{2},\mathbf{2},\mathbf{2}\right) \right)
\,\cong \,(\mathbf{3},\mathbf{1},\mathbf{1})^{\oplus 3}\,\oplus \,(\mathbf{1}%
,\mathbf{3},\mathbf{1})^{\oplus 3}\,\oplus \,(\mathbf{1},\mathbf{1},\mathbf{3%
})^{\oplus 3}\,\oplus \,\ldots ,  \label{decomppp}
\end{equation}%
where the dots denote other $25$ terms, which are not relevant for our
purposes. $\mathbf{3}$ denotes the adjoint (spin $s=1$) irrep. of $SL(2,%
\mathbb{R})$, which has a unique quadratic invariant (the $SL(2,\mathbb{R}%
)\sim SO(2,1)$ Cartan-Killing invariant metric $\eta =diag(1,1-1)$); as a
consequence, since $\mathbf{1}$ denotes the singlet, there is a unique
quadratic invariant induced onto the $(\mathbf{3},\mathbf{1},\mathbf{1})$, $(%
\mathbf{1},\mathbf{3},\mathbf{1})$ and $(\mathbf{1},\mathbf{1},\mathbf{3})$
of $SL\left( 2,\mathbb{R}\right) ^{3}$. Thus, from the representations in
the r.h.s. of (\ref{decomppp}), one obtains $3\cdot 3=9$ quadratic $SL\left(
2,\mathbb{R}\right) ^{3}$-invariant structures:%
\begin{equation}
\begin{array}{l}
\exists !\left( \mathbf{1},\mathbf{1},\mathbf{1}\right) \in (\mathbf{3},%
\mathbf{1},\mathbf{1})\otimes _{s}(\mathbf{3},\mathbf{1},\mathbf{1})~~(\text{%
3~times}); \\
\exists !\left( \mathbf{1},\mathbf{1},\mathbf{1}\right) \in (\mathbf{1},%
\mathbf{3},\mathbf{1})\otimes _{s}(\mathbf{1},\mathbf{3},\mathbf{1})~~(\text{%
3~times}); \\
\exists !\left( \mathbf{1},\mathbf{1},\mathbf{1}\right) \in (\mathbf{1},%
\mathbf{1},\mathbf{3})\otimes _{s}(\mathbf{1},\mathbf{1},\mathbf{3})~~(\text{%
3~times}).%
\end{array}%
\end{equation}

One can check that these $9$ invariants, together with $F_{0}$ (\ref{Inv}),
yield $10$ linearly independent invariants in $S_{4^{3}}\left( \left(
\mathbf{2},\mathbf{2},\mathbf{2}\right) \right) $. Thus, as announced, they
do provide a complete basis for the $10$-dimensional space of $\left(
SL_{h}\left( 3,\mathbb{R}\right) \times SL\left( 2,\mathbb{R}\right)
^{3}\right) $-invariant homogeneous polynomials of degree $12$ in the $(%
\mathbf{3},\mathbf{2},\mathbf{2},\mathbf{2})$, as resulting from Table (\ref%
{tab:stu}).

\subsection{\label{dec111}Explicit Construction}

Let $f_{1},f_{2},f_{3}$ and $e_{1},\ldots ,e_{8}$ be the standard basis of $%
\mathbb{R}^{3}=:\mathbf{3}$ of $SL_{h}\left( 3,\mathbb{R}\right) $, and of $(%
\mathbf{2},\mathbf{2},\mathbf{2})$ of $SL\left( 2,\mathbb{R}\right) ^{3}$,
respectively. Thus, any tensor $t\in (\mathbf{3},\mathbf{2},\mathbf{2},%
\mathbf{2})$ of $SL_{h}\left( 3,\mathbb{R}\right) \times SL\left( 2,\mathbb{R%
}\right) ^{\otimes 3}$ can be written as\footnote{%
In $3$-centered BH physics, the $c_{ij}$ ($i=1,2,3,j=1,...,8$) spanning the $%
(\mathbf{3},\mathbf{2},\mathbf{2},\mathbf{2})$ of $SL_{h}\left( 3,\mathbb{R}%
\right) \times SL\left( 2,\mathbb{R}\right) ^{3}$ would usually be denoted
as $Q_{a}^{M}$, with $a=1,2,3$ being the \textit{\textquotedblleft
horizontal"} $SL_{h}\left( 3,\mathbb{R}\right) $-index, and $M=1,...,8$
denoting the $U$-duality $SL\left( 2,\mathbb{R}\right) ^{3}$-index.} (using
the notation of Sec.\ (\ref{invsl}), and in particular denoting the
\textquotedblleft horizontal" index by $i=1,2,3$)%
\begin{equation}
t=\sum_{i=1,2,3,~j=1,...,8}c_{ij}f_{i}\otimes
e_{j}=\sum_{i=1}^{3}f_{i}\otimes r_{i},
\end{equation}%
for certain uniquely determined elements $r_{i}\in (\mathbf{2},\mathbf{2},%
\mathbf{2})$ of $SL\left( 2,\mathbb{R}\right) ^{3}$.

As discussed in Sec.\ \ref{geometric}, the \textit{Pl\"{u}cker coordinates} $%
p_{i_{1}i_{2}i_{3}}(t)$ of the tensor $t$ are the determinants of the $%
3\times 3$ matrices formed by the the rows $i_{1},i_{2},i_{3}$ of the $%
8\times 3$ matrix which has columns $r_{1},r_{2},r_{3}$:
\begin{equation}
p_{i_{1}i_{2}i_{3}}(t)\,=\,\det
\begin{pmatrix}
c_{1i_{1}} & c_{2i_{1}} & c_{3i_{1}} \\
c_{1i_{2}} & c_{2i_{2}} & c_{3i_{2}} \\
c_{1i_{3}} & c_{2i_{3}} & c_{3i_{3}}%
\end{pmatrix}%
.  \label{PC}
\end{equation}%
This is the formula defining $p_{i_{1}i_{2}i_{3}}=p_{\left[ i_{1}i_{2}i_{3}%
\right] }$, and their number is indeed $\binom{8}{3}=56$.

In the $stu$ model $G_{4}=SL\left( 2,\mathbb{R}\right) ^{3}$, with Lie
algebra $\mathfrak{G}_{4}=\mathfrak{sl}\left( 2,\mathbb{R}\right) ^{\oplus
3} $. Denoting by $X_{a}$ (raising operator), $Y_{a}$ (lowering operator),
and $H_{a}:=[X_{a},Y_{a}]$ the standard generators of the $a$-th ($a=1,2,3$)
copy of the Lie algebra $\mathfrak{sl}\left( 2,\mathbb{R}\right) $, the
action of $\mathfrak{sl}\left( 2,\mathbb{R}\right) ^{\oplus 3}$ on a vector $%
(c_{i1},\ldots ,c_{i8})\in (\mathbf{2},\mathbf{2},\mathbf{2})$ can be
realized through the identification ($i=1,2,3$) \footnote{%
In physics literature, the basis $\left\{ x_{\mathbf{abc}}\right\} _{\mathbf{%
a},\mathbf{b},\mathbf{c}=0,1}$ is named \textit{qubit basis}, because it
naturally occurs in the quantum entanglement of three \textit{qubits} in
\textit{Quantum Information Theory}. For relation to other symplectic frames
in the $stu$ model as well as recent developments related to the \textit{%
BH/qubit correspondence}, see \textit{e.g.} \cite{STU,KL-QI,BMOS-1,BFMY-FI}
and \cite{QI-Refs}, respectively.}
\begin{equation}
\begin{array}{l}
(c_{11},\ldots
,c_{18})\,=%
\,(x_{000},x_{001},x_{010},x_{011},x_{100},x_{101},x_{110},x_{111}); \\
(c_{21},\ldots
,c_{28})\,=%
\,(y_{000},y_{001},y_{010},y_{011},y_{100},y_{101},y_{110},y_{111}); \\
(c_{31},\ldots
,c_{38})\,=%
\,(z_{000},z_{001},z_{010},z_{011},z_{100},z_{101},z_{110},z_{111}),%
\end{array}
\label{identif-1}
\end{equation}%
where the fundamental (spin $s=1/2$) irrep. $\mathbf{2}$ of $SL\left( 2,%
\mathbb{R}\right) $ is spanned by the indices $\mathbf{a}=0,1$. For example,
the first copy of $\mathfrak{sl}\left( 2,\mathbb{R}\right) $ in $\mathfrak{sl%
}\left( 2,\mathbb{R}\right) ^{\oplus 3}$ acts on the $\mathbf{x}_{\mathbf{abc%
}}$ (equivalently denoting $x_{\mathbf{abc}}$ or $y_{\mathbf{abc}}$ or $z_{%
\mathbf{abc}}$) as follows:
\begin{equation}
X_{1}x_{\mathbf{abc}}\,=\,\left\{
\begin{array}{rl}
0 & \mbox{if}\;\mathbf{a}=0; \\
x_{0\mathbf{bc}} & \mbox{if}\;\mathbf{a}=1;%
\end{array}%
\right. \qquad Y_{1}x_{\mathbf{abc}}\,=\,\left\{
\begin{array}{rl}
x_{1\mathbf{bc}} & \mbox{if}\;\mathbf{a}=0; \\
0 & \mbox{if}\;\mathbf{a}=1;%
\end{array}%
\right. \qquad H_{1}x_{\mathbf{abc}}\,=\,\left\{
\begin{array}{rl}
x_{\mathbf{abc}} & \mbox{if}\;\mathbf{a}=0; \\
-x_{\mathbf{abc}} & \mbox{if}\;\mathbf{a}=1,%
\end{array}%
\right.
\end{equation}%
and similarly for the other two copies.

Then, one can compute the action of $\mathfrak{sl}\left( 2,\mathbb{R}\right)
^{\oplus 3}$ on the \textit{Pl\"{u}cker coordinates} (\ref{PC}), exploiting
the fact that elements of $\mathfrak{sl}\left( 2,\mathbb{R}\right) ^{\oplus
3}$ act as \textit{derivations} on the $p_{i_{1}i_{2}i_{3}}$'s themselves.
For example, by using the identification (\ref{identif-1}), the action of $%
X_{1}$ of the first copy of $\mathfrak{sl}\left( 2,\mathbb{R}\right) $ in $%
\mathfrak{sl}\left( 2,\mathbb{R}\right) ^{\oplus 3}$ on $p_{167}$ (\ref{PC})
reads
\begin{equation}
X_{1}p_{167}\,=\,X_{1}\det
\begin{pmatrix}
x_{000} & y_{000} & z_{000} \\
x_{101} & y_{101} & z_{101} \\
x_{110} & y_{110} & z_{110}%
\end{pmatrix}%
=\det
\begin{pmatrix}
x_{000} & y_{000} & z_{000} \\
x_{001} & y_{001} & z_{001} \\
x_{110} & y_{110} & z_{110}%
\end{pmatrix}%
+\det
\begin{pmatrix}
x_{000} & y_{000} & z_{000} \\
x_{101} & y_{101} & z_{101} \\
x_{010} & y_{010} & z_{010}%
\end{pmatrix}%
;
\end{equation}%
therefore, by using the antisymmetry of the \textit{Pl\"{u}cker coordinates}
(\ref{PC}), one finds that $X_{1}p_{167}=p_{127}-p_{136}$. In this way, one
can compute the action of each of the $9$ generators $\left\{
X_{1},Y_{1},H_{1},X_{2},Y_{2},H_{2},X_{3},Y_{3},H_{3}\right\} $ of $%
\mathfrak{sl}\left( 2,\mathbb{R}\right) ^{\oplus 3}$ on the representation $%
\wedge ^{3}(\mathbf{2},\mathbf{2},\mathbf{2})$ (realized in terms of \textit{%
Pl\"{u}cker coordinates} (\ref{PC}); also \textit{cfr.} (\ref{P-map})) of $%
SL\left( 2,\mathbb{R}\right) ^{3}$. Such an action then extends to an action
by \textit{derivations} on polynomials in the $p_{i_{1}i_{2}i_{3}}$'s
themselves.

\subsubsection{\label{dec112}The Representation $V(a_{1},a_{2},a_{3})$}

Let us now consider the realization of the representation $%
V(a_{1},a_{2},a_{3})$ of $G_{4}=SL\left( 2,\mathbb{R}\right) ^{3}$ on the
space of homogeneous polynomials; here, we use the standard notation in
which $V(a_{1},a_{2},a_{3}):=\left( \mathbf{a}_{1}\mathbf{+1},\mathbf{a}_{2}%
\mathbf{+1},\mathbf{a}_{3}\mathbf{+1}\right) $, and thus it has (real)
dimension $(a_{1}+1)(a_{2}+1)(a_{3}+1)$ (namely, $\left(
a_{1},a_{2},a_{3}\right) $ denote the weights of the vector space $V$ as $%
SL\left( 2,\mathbb{R}\right) ^{3}$-representation).

The \textit{highest weight} vector $v\in V(a_{1},a_{2},a_{3})$ satisfies
\begin{equation}
\left\{
\begin{array}{l}
H_{i}v\,=\,a_{i}v; \\
X_{i}v=0;%
\end{array}%
\right. \quad i\,=\,1,2,3.  \label{hw-1}
\end{equation}%
Thus, $V(a_{1},a_{2},a_{3})$ can be realized as the vector space spanned by
certain combinations of powers of lowering operators $X_{i}$'s on its
highest weight vector $v$ itself:
\begin{equation}
V(a_{1},a_{2},a_{3})\,=\,\langle Y_{1}^{k}Y_{2}^{l}Y_{3}^{m}v\;:\;0\leq
k\leq a_{1},\;0\leq l\leq a_{2},\;0\leq m\leq a_{3}\,\rangle .
\end{equation}%
By virtue of (\ref{hw-1}), the vector $Y_{1}^{k}Y_{2}^{l}Y_{3}^{m}v\in
V(a_{1},a_{2},a_{3})$ is again an eigenvector of all three $H_{i}$'s with
weight $(a_{1}-2k,a_{2}-2l,a_{3}-2m)$.\medskip

We are now going to exploit this general description\ in order to explicitly
construct the $10$ \linebreak $\left( SL_{h}\left( 3,\mathbb{R}\right)
\times SL\left( 2,\mathbb{R}\right) ^{3}\right) $-invariant homogeneous
polynomials of degree $12$ in the $(\mathbf{3},\mathbf{2},\mathbf{2},\mathbf{%
2})$ considered in Secs. \ref{oci} and \ref{oci-2}, which constitute a
complete basis for the corresponding $10$-dimensional vector space resulting
from Table (\ref{tab:stu}).


\subsubsection{\label{111}The $(\mathbf{2},\mathbf{2},\mathbf{2})$ in $%
S_{1^{3}}((\mathbf{2},\mathbf{2},\mathbf{2}))$}

Below (\ref{s111}), we observed that there is a (unique) irreducible
tri-fundamental $SL\left( 2,\mathbb{R}\right) ^{3}$-representation $%
V(1,1,1)=:(\mathbf{2},\mathbf{2},\mathbf{2})$ in $S_{1^{3}}((\mathbf{2},%
\mathbf{2},\mathbf{2}))=\wedge ^{3}(\mathbf{2},\mathbf{2},\mathbf{2})=:(%
\mathbf{2},\mathbf{2},\mathbf{2})_{a}^{\otimes 3}$. In order to characterize
it, we here determine its highest weight vector.

Besides $(\mathbf{2},\mathbf{2},\mathbf{2})$, also each of the other $3$
irreducible summands of $S_{1^{3}}((\mathbf{2},\mathbf{2},\mathbf{2}))$ in
the r.h.s. of (\ref{s111}) has a vector with weight $(1,1,1)$, therefore the
weight space $S_{1^{3}}((\mathbf{2},\mathbf{2},\mathbf{2}))_{(1,1,1)}$ is
four-dimensional:%
\begin{eqnarray}
S_{1^{3}}((\mathbf{2},\mathbf{2},\mathbf{2}))_{(1,1,1)} &:&=\{v\in
\,S_{1^{3}}((\mathbf{2},\mathbf{2},\mathbf{2})):\;H_{i}v\,=\,v,\quad
i=1,2,3\,\}  \notag \\
&=&\langle p_{145},\,p_{136},\,p_{235},\,p_{127}\,\rangle ,
\end{eqnarray}%
{\ } as one can check within the conventions adopted above. The unique (up
to a scalar multiple) highest weight vector in this space is
\begin{equation}
v:=\,p_{145}-p_{136}-p_{127},\qquad \mbox{so}\quad \langle v\rangle
\,=\,\cap _{i=1}^{3}\ker (X_{i})\cap S_{1^{3}}((\mathbf{2},\mathbf{2},%
\mathbf{2}))_{(1,1,1)}.  \label{hw-111}
\end{equation}%
Thus, an isomorphism between $(\mathbf{2},\mathbf{2},\mathbf{2})\subset
S_{1^{3}}((\mathbf{2},\mathbf{2},\mathbf{2}))$ and $(\mathbf{2},\mathbf{2},%
\mathbf{2})$ itself can be obtained, by setting
\begin{equation}
x_{klm}\,:=\,Y_{1}^{k}Y_{2}^{l}Y_{3}^{m}v,\qquad k,l,m\,\in \,\{0,1\},
\end{equation}%
where $v$ is defined in (\ref{hw-111}).

The usual expression of the $SL\left( 2,\mathbb{R}\right) ^{3}$-invariant
\textit{Cayley's hyperdeterminant} $\mathcal{I}_{4}$ \cite{Cayley} in the
tri-fundamental $(\mathbf{2},\mathbf{2},\mathbf{2})$ as a quartic
homogeneous polynomial in the $x_{ijk}$'s \cite{Duff-Cayley} (in \textit{%
qubit basis}; \textit{cfr.} footnote 30) produces a degree-$4$ polynomial in
the \textit{Pl\"{u}cker coordinates} $p_{ijk}(t)$ (\ref{PC}). As a
polynomial in the $c_{ij}$ (\textit{cfr. e.g.} the first line of (\ref%
{identif-1})), such a polynomial is then $\left( SL_{h}\left( 3,\mathbb{R}%
\right) \times SL\left( 2,\mathbb{R}\right) ^{3}\right) $-invariant
homogeneous of degree $12$ in the $(\mathbf{3},\mathbf{2},\mathbf{2},\mathbf{%
2})$; indeed, as expected, one can check that it coincides with the
invariant $F_{0}(t)$ (\ref{Inv}).


\subsubsection{\label{002} The $(\mathbf{1},\mathbf{1},\mathbf{3})^{\oplus
3} $ in $S_{2^{3}}\left( \left( \mathbf{2},\mathbf{2},\mathbf{2}\right)
\right) $}

The $SL\left( 2,\mathbb{R}\right) ^{3}$-representation $S_{2^{3}}\left(
\left( \mathbf{2},\mathbf{2},\mathbf{2}\right) \right) $ is a
sub-representation of $S^{2}(S_{1^{3}}\left( \left( \mathbf{2},\mathbf{2},%
\mathbf{2}\right) \right) )=:\left( (\mathbf{2},\mathbf{2},\mathbf{2}%
)_{a}^{\otimes 3}\right) _{s}^{\otimes 2}$, which is the space of homogenous
polynomials of degree $2$ in the \textit{Pl\"{u}cker coordinates} $%
p_{ijk}(t) $; in fact, by substituting the cubic polynomials (\ref{PC}) in
the $c_{ij}$ for these $p_{ijk}$, one gets a vector space of degree-$6$
homogeneous polynomials in the $c_{ij}$'s, which is nothing but $%
S_{2^{3}}\left( \left( \mathbf{2},\mathbf{2},\mathbf{2}\right) \right) $.

Using this fact, one can first determine the weight space%
\begin{eqnarray}
S^{2}(S_{1^{3}}\left( \left( \mathbf{2},\mathbf{2},\mathbf{2}\right) \right)
)_{(0,0,2)} &:&=\{v\in \,S^{2}(S_{1^{3}}\left( \left( \mathbf{2},\mathbf{2},%
\mathbf{2}\right) \right) ):\;H_{i}v\,=\,0,\quad i=1,2,\quad
H_{3}v\,=\,2v\,\}  \notag \\
&=&\langle p_{168}p_{137},\ldots \rangle ,  \label{vvv}
\end{eqnarray}%
{\ } which has dimension $52$.

Next, by computing the images of the $52$ basis elements under the raising
operators $X_{i}$, $i=1,2,3$, one finds the highest weight vectors in $%
S^{2}(S_{1^{3}}\left( \left( \mathbf{2},\mathbf{2},\mathbf{2}\right) \right)
)$ (\ref{vvv}), which result to span a $5$-dimensional sub-space of such a
weight space. As they are rather complicated (and not particularly
illuminating) homogeneous polynomials of degree $2$ in the \textit{Pl\"{u}%
cker coordinates} $p_{ijk}(t)$, we will refrain from reporting them here
explicitly.

Then, by recalling (\ref{PC}), one can express $p_{ijk}$ in terms of the
coordinates $c_{ij}$, thus obtaining a $3$-dimensional sub-space. Let $%
\left\{ g_{1},g_{2},g_{3}\right\} $ be a basis of this sub-space; therefore,
$(\mathbf{1},\mathbf{1},\mathbf{3})^{\oplus 3}\subset S_{2^{3}}\left( \left(
\mathbf{2},\mathbf{2},\mathbf{2}\right) \right) $ (\textit{cfr.} (\ref%
{decomppp})) is spanned by $\left\{
g_{i},Y_{3}(g_{i}),Y_{3}^{2}(g_{i})\right\} _{i=1,2,3}$. The $SL\left( 2,%
\mathbb{R}\right) ^{3}$-representation $V(0,0,2)=:(\mathbf{1},\mathbf{1},%
\mathbf{3})$ has a unique invariant in $S^{2}V(0,0,2)=:(\mathbf{1},\mathbf{1}%
,\mathbf{3})_{s}^{\otimes 2}$, given by the Cartan-Killing metric in the
adjoint (spin $s=1$) irrep. $\mathbf{3}$ of the third copy of $SL\left( 2,%
\mathbb{R}\right) $ in $SL\left( 2,\mathbb{R}\right) ^{3}$ itself, and whose
expression in terms of $\left\{ g,Y_{3}(g),Y_{3}^{2}(g)\right\} $ is given
by
\begin{equation}
2gY_{3}^{2}(g)-\left( Y_{3}(g)\right) ^{2}.
\end{equation}

Thus, in $S^{2}\left( S_{2^{3}}\left( \left( \mathbf{2},\mathbf{2},\mathbf{2}%
\right) \right) \right) \subset S_{4^{3}}\left( \left( \mathbf{2},\mathbf{2},%
\mathbf{2}\right) \right) $, one gets $3$ $\left( SL_{h}\left( 3,\mathbb{R}%
\right) \times SL\left( 2,\mathbb{R}\right) ^{3}\right) $-invariant
homogeneous polynomials of degree $12$ in the $(\mathbf{3},\mathbf{2},%
\mathbf{2},\mathbf{2})$, which can be checked to be linearly independent as
polynomials in the $c_{ij}$'s.

By considering also the results of the same procedure repeated for $(\mathbf{%
3},\mathbf{1},\mathbf{1})^{\oplus 3}\subset S_{2^{3}}\left( \left( \mathbf{2}%
,\mathbf{2},\mathbf{2}\right) \right) $ as well as for $(\mathbf{1},\mathbf{3%
},\mathbf{1})^{\oplus 3}\subset S_{2^{3}}\left( \left( \mathbf{2},\mathbf{2},%
\mathbf{2}\right) \right) $ (\textit{cfr.} (\ref{decomppp})), one obtains a
total of $3$ $SL\left( 2,\mathbb{R}\right) ^{3}$-invariant homogeneous
polynomials of degree $12$ in $S^{2}\left( S_{2^{3}}\left( \left( \mathbf{2},%
\mathbf{2},\mathbf{2}\right) \right) \right) \subset S_{4^{3}}\left( \left(
\mathbf{2},\mathbf{2},\mathbf{2}\right) \right) $.


\subsubsection{\label{co10in}$stu$ \textit{Triality}}

In order to determine the remaining relevant $6$ invariants of degree $12$,
one can now use the action of the symmetric group $S_{3}$ on $R=\left(
\mathbf{2},\mathbf{2},\mathbf{2}\right) $ by permuting the tensor
components, so $(12)\in S_{3}$ will map $x_{abc}$ to $x_{bac}$, \textit{etc.}%
. Consequently, $S_{3}$ will also act on the $c_{ij}$'s, as well as on the Pl%
\"{u}cker coordinates $p_{ijk}$. As we will see below, in the context of $%
stu $ black holes, the invariance under $S_{3}$ must be enforced, because it
corresponds to the \textit{triality symmetry} \cite{STU} exhibited by such a
model of $\mathcal{N}=2$, $D=4$ supergravity.

Using this action, the $3$ invariants just found in\ Sec.\ \ref{002} give
rise to the required set of $9$ invariants.

Including the invariant from Sec.\ \ref{oci} (which, as mentioned above,
matches the one obtained in Sec.\ \ref{111}), one gets a total of $10$
invariants of degree $12$ in the $c_{ij}$'s.

Thus, we constructed a basis $\left\{ \mathbf{I}_{12,\alpha }\right\}
_{\alpha =1,...,10}$ for the $10$-dimensional vector space of \linebreak $%
\left( SL_{h}\left( 3,\mathbb{R}\right) \times SL\left( 2,\mathbb{R}\right)
^{3}\right) $-invariant homogeneous polynomials of degree $12$ in the $(%
\mathbf{3},\mathbf{2},\mathbf{2},\mathbf{2})$ (resulting from Table (\ref%
{tab:stu})).\medskip

As degree-$12$ homogeneous polynomials in the $c_{ij}\in \left( \mathbf{3},%
\mathbf{2},\mathbf{2},\mathbf{2}\right) $ of $SL_{h}\left( 3,\mathbb{R}%
\right) \times SL\left( 2,\mathbb{R}\right) ^{3}$ (realized \textit{e.g.}
through the identification (\ref{identif-1})), they have far too many terms,
rendering their explicit expression cumbersome and not particularly
illuminating. However, we observe that each of the invariants $\mathbf{I}%
_{12,\alpha }$ can be rewritten as%
\begin{equation}
\mathbf{I}_{12,\alpha }=\sum_{\beta =1}^{10}C_{\alpha \beta }M_{\beta }+...,
\label{decc}
\end{equation}%
where $C_{\alpha \beta }\in \mathbb{Z}$, and $M_{\beta }$ ($\beta =1,\ldots
,10$) denotes the following set of monomials:%
\begin{eqnarray}
M_{1}
&:&=c_{18}^{4}c_{23}^{2}c_{25}c_{26}c_{31}^{3}c_{32},~~~M_{2}:=c_{18}^{4}c_{23}^{2}c_{25}^{2}c_{31}^{2}c_{32}^{2},~~~M_{3}:=c_{18}^{4}c_{22}c_{24}c_{25}^{2}c_{31}^{3}c_{33},
\notag \\
M_{4}
&:&=c_{18}^{4}c_{22}c_{23}c_{26}c_{27}c_{31}^{4},~~~M_{5}:=c_{18}^{4}c_{22}c_{23}c_{25}c_{27}c_{31}^{3}c_{32},~~M_{6}:=c_{18}^{4}c_{22}^{2}c_{27}^{2}c_{31}^{4},
\notag \\
M_{7}
&:&=c_{11}^{3}c_{16}c_{22}^{2}c_{27}^{2}c_{33}c_{38}^{3},~~~~M_{8}:=c_{11}^{3}c_{14}c_{22}^{2}c_{27}^{2}c_{35}c_{38}^{3},~~~M_{9}:=c_{11}^{3}c_{14}c_{22}c_{23}c_{26}c_{27}c_{35}c_{38}^{3},
\notag \\
M_{10} &:&=c_{11}^{3}c_{14}c_{23}^{2}c_{26}^{2}c_{35}c_{38}^{3},
\label{set-M}
\end{eqnarray}%
which completely characterize each $\mathbf{I}_{12,\alpha }$. Indeed, the
dots in the right-hand side of (\ref{decc}) stand for many other linear
combinations of monomials which are linearly independent on the $M_{\beta }$%
's, but which can be determined by the action of the whole group $%
SL_{h}\left( 3,\mathbb{R}\right) \times SL\left( 2,\mathbb{R}\right) ^{3}$
itself, once the $C_{\alpha \beta }$'s are specified. {\ } Of course, there
are many sets of $10$ monomials with the property (\ref{decc}). Given the
set $M_{\beta }$ (\ref{set-M}), each element $\mathbf{I}_{12,\alpha }$ of
the $10$-dimensional complete basis $\left\{ \mathbf{I}_{12,\alpha }\right\}
_{\alpha =1,...,10}$ constructed above can be written as a vector in $%
\mathbb{Z}^{10}$.

Let us make some examples.

The invariant $F_{0}$ (\ref{Inv}), constructed in Sec.\ \ref{oci} as well as
in Sec.\ \ref{111}, has coordinates\footnote{%
What we actually write in (\ref{G_0}), (\ref{G_0-post}) and (\ref{G_0-post-2}%
) are $11$ integers ($a_{1},...,a_{11}$) such that%
\begin{equation*}
\sum_{\alpha =1}^{10}a_{\alpha }\mathbf{I}_{12,\alpha }+a_{11}F_{0}=0.
\end{equation*}%
}
\begin{equation}
F_{0}\equiv (0,0,0,-2,4,1,4,4,-8,4,4).  \label{G_0}
\end{equation}

On the other hand, the following three vectors correspond to the
aforementioned basis $\left\{ g_{1},g_{2},g_{3}\right\} $ the $3$%
-dimensional sub-space of invariants obtained from $(\mathbf{1},\mathbf{1},%
\mathbf{3})^{\oplus 3}\subset S_{2^{3}}\left( \left( \mathbf{2},\mathbf{2},%
\mathbf{2}\right) \right) $ (\textit{cfr.} Sec.\ \ref{002}):{\
\begin{equation}
\begin{array}{l}
g_{1}\equiv (-6,6,1,0,0,0,0,0,9,-9,-9); \\
g_{2}\equiv (2,-2,-1,2,-4,-2,10,10,-15,5,5); \\
g_{3}\equiv (2,-1,0,2,-2,-1,6,4,-8,4,4).%
\end{array}%
\quad  \label{G_0-post}
\end{equation}%
}

As mentioned above, the $stu$ \textit{triality }symmetry (implemented as the
symmetric group $S_{3}$) permutes the subspaces $(\mathbf{1},\mathbf{1},%
\mathbf{3})^{\oplus 3}$, $(\mathbf{1},\mathbf{3},\mathbf{1})^{\oplus 3}$ and
$(\mathbf{3},\mathbf{1},\mathbf{1})^{\oplus 3}$; as a consequence, there is
a $3$-dimensional sub-space of \textit{triality-invariant} $\left(
SL_{h}\left( 3,\mathbb{R}\right) \times SL\left( 2,\mathbb{R}\right)
^{3}\right) $-invariants in their direct sum,\footnote{%
Its complement is the direct sum of three $2$-dimensional (irreducible) $%
S_{3}$-representations.} which is the space spanned by the vectors%
\begin{equation}
\begin{array}{l}
F_{1}\equiv (1,-1,1,1,-2,-1,3,3,-14,3,3); \\
F_{2}\equiv (-5,5,-5,7,-14,-1,-7,-7,-10,-7,-7); \\
F_{3}\equiv (4,-3,0,4,-6,-3,4,2,-22,8,8).%
\end{array}%
~  \label{G_0-post-2}
\end{equation}

By adding the invariant $F_{0}$ (\ref{Inv}) (or equivalently, through (\ref%
{decc})-(\ref{set-M}), (\ref{G_0})), which is also \textit{triality-invariant%
}, out of $\left\{ \mathbf{I}_{12,\alpha }\right\} _{\alpha =1,...,10}$ one
gets a $4$-dimensional basis $\left\{ F_{0},F_{1},F_{2},F_{3}\right\} $ for
degree-$12$ homogeneous polynomials invariant under the action of $%
S_{3}\times SL_{h}\left( 3,\mathbb{R}\right) \times SL\left( 2,\mathbb{R}%
\right) ^{3}$ on $\left( \mathbf{3},\mathbf{2},\mathbf{2},\mathbf{2}\right) $%
. Thus, as anticipated in Sec.\ \ref{stu}, $4$ is the (real) dimension of
the vector space of degree-$12$ $\left( SL_{h}\left( 3,\mathbb{R}\right)
\times SL\left( 2,\mathbb{R}\right) ^{3}\right) $-invariant polynomials
relevant for the $3$-centered BHs in the $\mathcal{N}=2$, $D=4$ $stu$ model.

\section*{Acknowledgements}

We would like to thank Sergio Ferrara for useful discussions and
correspondence.

A. M. would also like to thank Giuseppe Torri for interesting discussions on
the Hilbert series and plethystic logarithm.


\end{document}